\documentclass[12pt,a4paper]{article}
\usepackage{jheppub_modified}

\usepackage{subfigure}
\usepackage{ulem}

\title{
Imaging black holes through AdS/CFT
}

\author[1]{Koji Hashimoto}
\author[2]{Shunichiro Kinoshita}
\author[1,3]{Keiju Murata}

\affiliation[1]{Department of Physics, Osaka University, Toyonaka, Osaka 560-0043, Japan}
\affiliation[2]{Department of Physics, Chuo University, Tokyo 112-8551, Japan}
\affiliation[3]{Department of Physics, College of Humanities and Sciences, Nihon University, Tokyo 156-8550, Japan}

\abstract{%
Thermal states in some quantum field theories (QFTs) correspond to black holes in asymptotically AdS spacetime in the AdS/CFT correspondence.
We propose a direct procedure  
to construct holographic images of the black hole in the bulk from a
given response function of the QFT on the boundary.
The response function with respect to an external source 
corresponds to the asymptotic data of the bulk field generated by the source on the
AdS boundary.
According to the wave optics, we can obtain the images from the bulk field propagating in the bulk spacetime.
For a thermal state on two-dimensional sphere dual to
Schwarzschild-AdS$_4$ black hole, 
we demonstrate that the holographic
images gravitationally lensed by the black hole can be constructed from
the response function.
In particular, the Einstein rings on the image can be clearly observed
and their radius depends on the total energy of the QFT thermal state.
These results are consistent with the size of the photon sphere of the
black hole calculated in geometrical optics.
This implies that, if there exists a dual gravitational picture for a
given quantum system, we would be able to probe existence of the dual
black hole by the Einstein rings constructed from observables of the quantum system.
}
\preprint{OU-HET-982}

\begin{document}
\maketitle

\section{Introduction}
One of the definite goals of the research of the holographic principle,
or the AdS/CFT correspondence \cite{Maldacena:1997re,Gubser:1998bc,Witten:1998qj},
is to find what class of quantum field theories (QFTs) or quantum materials possesses 
their gravity dual. 
Is there any direct test for the existence of a gravity dual for a given material?

\begin{figure}
\begin{center}
\includegraphics[scale=0.8]{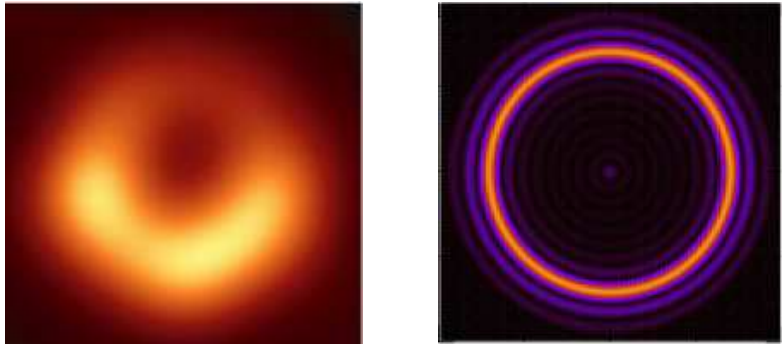}
\end{center}
 \caption{
(Left) Image of the black hole in M87 (This figure is taken from Ref.~\cite{EHT}.)
(Right) Image of the AdS black hole constructed from the response function.
 }
 \label{EHTfig}
\end{figure}

Among various gravitational physics, one of the most peculiar astrophysical 
objects is the black hole.
Gravitational lensing is one of fundamental phenomena by strong
gravity.
Let us consider that there is a light source behind a gravitational
body.
When the light source, the gravitational body, and observers are in
alignment, the observers will see a ring-like image of the light source,
i.e., the so-called Einstein ring.
If the gravitational body is a black hole, some light rays are so strongly
bended that they can go around the black hole many times, and especially
infinite times on the photon sphere.
As a result, multiple Einstein rings corresponding to winding numbers of
the light ray orbits emerge and infinitely concentrate
on the photon sphere.
Recently, the Event Horizon Telescope (EHT)~\cite{EHT}, which is an observational
project for imaging black holes, has captured the first image of the 
supermassive black hole in M87. (See the left panel of Fig.~\ref{EHTfig}.)
The dark area inside the photon sphere, in which light rays have been
absorbed by the black hole, is called black hole shadow~\cite{Falcke:1999pj}.
In  this paper, we propose a direct method to check the existence of a gravity dual from measurements in a given thermal QFT 
--- {\it imaging the dual black hole as an ``Einstein ring''}.

\begin{figure}
\begin{center}
\includegraphics[scale=0.6]{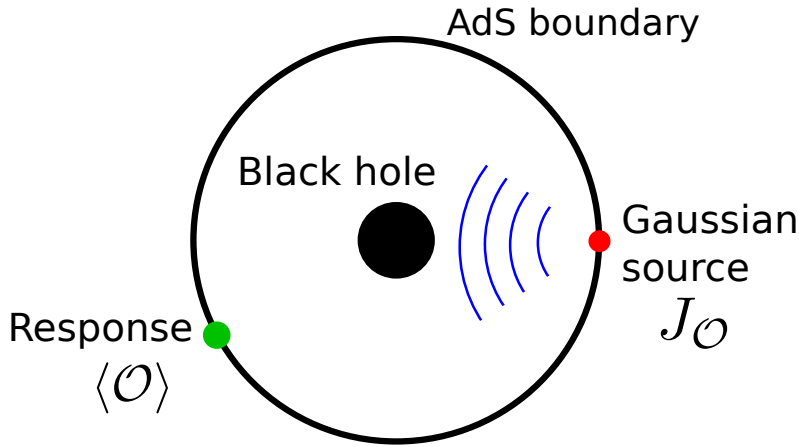}
\end{center}
\caption{
Our setup for imaging a dual black hole, the Schwarzschild-AdS$_4$ spacetime. 
An oscillating Gaussian source 
$J_\mathcal{O}$
is applied at a point on the AdS boundary.
Its response $\langle {\cal O}(x)\rangle$ is observed at another point on the boundary.
}
\label{setupfig}
\end{figure}

We demonstrate explicitly construction of holographic ``images'' of the dual black hole 
from the response function of the boundary QFT with external sources, as follows.
As the simplest example, we consider a $(2+1)$-dimensional 
boundary conformal field theory on a 2-sphere $S^2$ at a finite temperature,
and study 
a one-point function of a scalar operator ${\cal O}$ with its conformal dimension $\Delta_{\cal O} = 3$, 
under a time-dependent localized source $J_{\cal O}$.
The gravity dual is a black hole in the global AdS$_4$ and a probe
massless bulk scalar field in the spacetime.
The schematic picture of our setup is shown in Fig.~\ref{setupfig}.
The source $J_{\cal O}$, for which we employ a time-periodic localized Gaussian source with the frequency $\omega$, 
amounts to an AdS boundary condition for the scalar field.
Due to the time periodic boundary condition, 
a bulk scalar wave is injected into the bulk from the AdS boundary.
The scalar wave propagates inside the black hole spacetime 
and reaches other points on the $S^2$ of the AdS boundary.
We measure the local response function $e^{-i\omega t}\langle {\cal O}(\vec{x})\rangle$ which 
has the information about the bulk geometry of the black hole spacetime.

Using a wave-optical method, 
we find a formula which converts the response function $\langle {\cal O}(\vec{x})\rangle$ to the image 
of the dual black hole $|\Psi_\mathrm{S}(\vec{x}_\mathrm{S})|^2$ on a virtual screen:
\begin{equation}
 \Psi_\mathrm{S}(\vec{x}_\mathrm{S})= \int_{|\vec{x}|<d}d^2x\, \langle {\cal O}(\vec{x})\rangle e^{-\frac{i\omega}{f}\vec{x}\cdot\vec{x}_\mathrm{S}}\ ,
\label{lenstrans0}
\end{equation}
where $\vec{x}=(x,y)$ and $\vec{x}_\textrm{S}=(x_\textrm{S},y_\textrm{S})$ are Cartesian-like coordinates on boundary $S^2$ and 
the virtual screen, respectively, 
and we have set the origin of the coordinates to an observation point.
This operation is mathematically 
a Fourier transformation of the
response function on a small patch with the radius $d$
around the observation point.
Note that $f$ describes magnification of the image on the screen.
In wave optics, we have virtually used a lens with the focal
length $f$ and the radius $d$ to form the image.
The right panel of Fig.~\ref{EHTfig} shows a typical image of the AdS black hole computed
from the response function through our method.
The AdS/CFT calculation clearly gives a ring similar to the
observed Einstein ring.
Equation~(\ref{lenstrans0}) can be regarded as the dual quantity of Einstein ring caused by the
black hole in a thermal QFT.

Several criteria for QFTs to have a gravity dual have been proposed in some previous works.
For example, in a conformal field theory,
the existence of a planer expansion and
a large gap in the spectrum of anomalous dimensions 
has been conjectured to be the criterion for the existence of a gravity dual~\cite{Heemskerk:2009pn}.
There is also recent progress on emergent gravity based on quantum entanglement~\cite{Faulkner:2013ica},
although the entanglement entropy itself is not a physical observable.
The strong redshift of the black hole 
has been also used as the condition for the existence of gravity 
dual~\cite{Shenker:2013pqa,Maldacena:2015waa,Kitaev-talk}.
Our approach, checking the dual black hole by its image, gives an alternative.
The method is simple and can be applied to any QFT on a sphere, thus probing efficiently a black hole
of its possible gravity dual. 
Once we have a strongly correlated material on $S^2$, we can apply 
a localized external source such as electromagnetic waves and measure its response in principle.
Then, from Eq.~(\ref{lenstrans0}), 
we would be able to construct the image of the dual black hole if it exists.
The holographic image of black holes in a material, if observed by a tabletop experiment, may serve as a
novel entrance to the world of quantum gravity.

The organization of this paper is as follows. In Section~\ref{Sec:scalar}, we review
how to obtain the response function under a source in AdS/CFT, from a scalar field dynamics
in Schwarzschild-AdS geometry, and describe our time-oscillatory source in the boundary QFT.
In Section~\ref{imgwave}, 
we review image formation in wave optics. 
This leads us to our main formula for the dual quantity of the Einstein ring~\eqref{lenstrans0}.
In Section~\ref{ngeo}, we discuss null geodesics in the
Schwarzschild-AdS on the basis of geometrical optics so that we will
gain intuitive insight into our holographic images. 
Then in Section~\ref{ImAdSBH}, we provide our concrete images for the AdS black holes
seen from the QFT with the formula~\eqref{lenstrans0}. 
They show clear 
holographic images of black holes 
as
well as the Einstein rings. In Section~\ref{RingAngle}, we evaluate the Einstein radius (the size of the rings) in the images,
and provide its consistent understanding by geodesics. Then in Section~\ref{Analy}, 
we provide analytic examples of the pure AdS$_4$ and the BTZ black hole, for a comparison.
In Section~\ref{EringGreen}, we describe the image of the Einstein ring from holography in terms of the retarded Green function. We see that 
quasi-normal mode frequencies in the gravity side give major contribution for the formation of the Einstein ring.
Finally, Section~\ref{Concl} is for our conclusion and discussions.
Appendix~\ref{sec:detail} describes detailed numerical evaluation of solutions of the scalar field in the bulk.
Appendix~\ref{Val} is for the validity of the approximation of the geometric optics used in Section~\ref{ngeo}.

\section{Scalar field in Schwarzschild-AdS$_4$ spacetime}
\label{Sec:scalar}

We consider Schwarzschild-AdS$_4$ (Sch-AdS$_4$) with the spherical horizon: 
\begin{align}
ds^2=-F(r)dt^2+\frac{dr^2}{F(r)}+r^2(d\theta^2+\sin^2\theta d\varphi^2)\ ,\quad
F(r)=\frac{r^2}{R^2}+1-\frac{2 G M}{r}\ ,
\label{SchAdS4}
\end{align}
where $R$ is the AdS curvature radius 
(i.e., a negative cosmological constant $\Lambda = -3/R^2$) 
and $G$ is the Newton constant.
Since this is a black hole solution in the global AdS$_4$,
we consider a CFT on $\mathbf{R}_t \times S^2$ as the dual field theory.
The event horizon is located at $r=r_\mathrm{h}$ defined by $F(r_\mathrm{h})=0$
\footnote{
Although the Sch-AdS$_4$ with $r_\mathrm{h} \lesssim R$ is unstable in the canonical ensemble, 
it can be stable in the microcanonical ensemble~\cite{Hawking:1982dh}.  
We will consider black holes with $r_\mathrm{h} \lesssim R$ as well as those with $r_\mathrm{h} \gtrsim R$ in this paper.
}.
Using the horizon radius, 
the 
mass $M$ is written as
\begin{equation}
M=\frac{r_\mathrm{h}(r_\mathrm{h}^2+R^2)}{GR^2}\ .
\label{Mass}
\end{equation}
This corresponds to the total energy of the dual CFT.
In what follows, 
for simplicity and without loss of generality 
we take the unit of $R=1$.

We focus on dynamics of a massless scalar field in the Sch-AdS$_4$.
The Klein-Gordon equation for the massless scalar field, 
$\Box \Phi(t,r,\theta,\varphi)=0$, is written as
\begin{equation}
 -\frac{1}{F}\partial_t^2 \Phi+F\partial_r^2
  \Phi+\frac{(r^2F)'}{r^2}\partial_r \Phi+\frac{1}{r^2} D^2 \Phi=0\ ,
\label{scalareq0}
\end{equation}
where 
the prime denotes $r$-derivative and 
$D^2$ is the scalar Laplacian on unit $S^2$.

Near the AdS boundary ($r=\infty$), 
the asymptotic solution of the scalar field becomes
\begin{align}
 \Phi(t,r,\theta,\varphi)
=J_\mathcal{O}(t,\theta,\varphi) 
-\frac{1}{2r^2}(\partial_t^2 - D^2)J_\mathcal{O}(t,\theta,\varphi) 
+\frac{\langle \mathcal{O}(t,\theta,\varphi)\rangle }{r^3} +\cdots \ .
\label{phiinf0}
\end{align}
In the asymptotic expansion, we have two independent functions, $J_\mathcal{O}$ and $\langle \mathcal{O}\rangle$,  
which depend on boundary coordinates $(t,\theta,\varphi)$.
According to the AdS/CFT dictionary, the leading term
$J_\mathcal{O}$ and the subleading term $\langle \mathcal{O}\rangle$
correspond to the external scalar source and its response function in
the dual CFT, respectively~\cite{Klebanov:1999tb}. 

We consider that an axisymmetric and monochromatically oscillating Gaussian source is
localized at the south pole ($\theta=\pi$) of the boundary $S^2$:
\begin{equation}
 J_\mathcal{O}(t,\theta,\varphi)=e^{-i\omega t} g(\theta)\ ,\quad g(\theta)=\frac{1}{2\pi \sigma^2}\exp\left[-\frac{(\pi-\theta)^2}{2\sigma^2}\right]\ .
\label{BC}
\end{equation}
Note that 
we will ignore a tiny value of
the Gaussian tail at the north pole because we suppose $\sigma\ll \pi$.
In the bulk point of view, this source $J_\mathcal{O}$
determines the boundary condition of the scalar field at the AdS boundary.
We also impose the ingoing boundary condition on the horizon of the Sch-AdS$_4$. 
A schematic picture of our setup is shown in Fig.~\ref{setupfig}.
The scalar wave generated at the south pole of the AdS boundary 
propagates inside the bulk black hole spacetime 
and reaches the other points at the AdS boundary.
Imposing these boundary conditions, we solve Eq.~(\ref{scalareq0}) numerically and 
determine the solution of the scalar field in the bulk.
We summarize the detailed method to solve the Klein-Gordon equation in
Appendix \ref{sec:detail}.
We read off the response from the coefficient of $1/r^3$ in Eq.~(\ref{phiinf0}).
Because of the properties 
of the source~(\ref{BC}), the response function does not depend on
$\varphi$ and its time dependence is just given by $e^{-i\omega t}$.
Hence, we can write the response function as
\begin{equation}
 \langle \mathcal{O}(t,\theta,\varphi)\rangle=e^{-i\omega t}\langle \mathcal{O}(\theta)\rangle\ .
\end{equation}
\begin{figure}
\begin{center}
\includegraphics[scale=0.5]{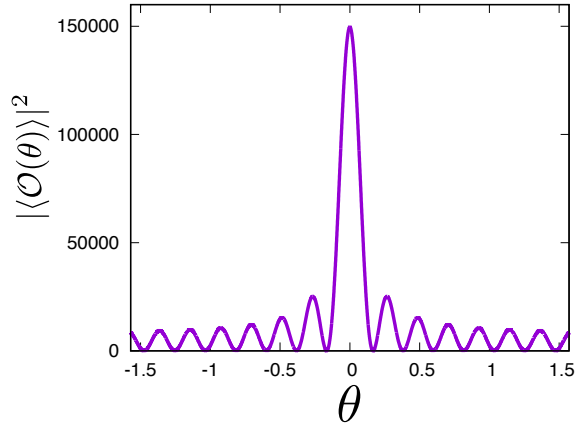}
\end{center}
\caption{
Absolute square of the response for $r_\mathrm{h}=0.3$, $\omega=20$ and $\sigma=0.2$.
}
\label{vev_abs}
\end{figure}
As an example, Fig.~\ref{vev_abs} shows the absolute square of the response function for $r_\mathrm{h}=0.3$, $\omega=20$ and $\sigma=0.2$.
From this response function itself, we cannot directly find
``black-hole-like images'' as we desired.
We have observed only the interference pattern resulting from the
diffraction of the scalar wave by the black hole.
In order to obtain images, we need to look at the response
function through a certain kind of optical system with a convex lens as
we will see in the next section.

\section{Image formation in wave optics}
\label{imgwave}

\begin{figure}
  \centering
  \subfigure[Transmitted wave]
 {\includegraphics[scale=1.3]{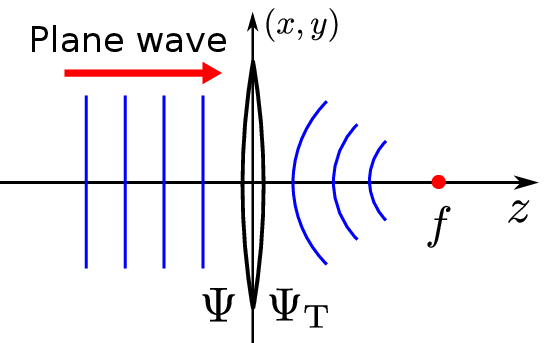}\label{lens_transmit}
  }
\subfigure[Image formation on the screen]
 {\includegraphics[scale=1.1]{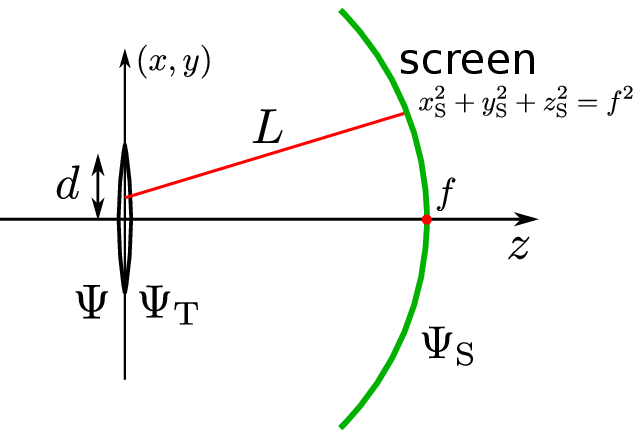}\label{imgf}
  }
 \caption{
(a) The lens converts plane waves into spherical waves and vice versa.
(b) The screen is located at $\{(x_\mathrm{S},y_\mathrm{S},z_\mathrm{S})|x_\mathrm{S}^2+y_\mathrm{S}^2+z_\mathrm{S}^2=f^2\}$.
 }
 \label{lensfig}
\end{figure}

We introduce an optical system consisting of a convex
lens and a spherical screen so that we will construct images of the
black hole from the response function on the basis of the wave optics~\cite{Optics}.
(See also Refs.~\cite{Nambu:2012wa,Kanai:2013rga,Nambu:2015aea}.) 
What role the convex lens plays in the wave optics is as follows.
The lens is regarded as a ``converter'' between plane and spherical waves as in Fig.~\ref{lens_transmit}.
The lens with focal length $f$ is located at $z=0$ 
and its focus is at $z=\pm f$.
We assume that 
the size of the lens is much smaller than the focal length $f$ and 
the lens is infinitely thin.
Imagine that a plane wave is irradiated to the lens from the left hand
side as shown in the figure.
Such a plane wave is converted into (a part of) spherical wave and it
converges at the focus located at $z=f$.
Inversely, if we consider a spherical wave emitted from the focus, it
should be converted into the plane wave in $z<0$.
Let $\Psi$ and $\Psi_\mathrm{T}$ denote the incident wave and the
transmitted wave, respectively.
Then, mathematically, the role of the convex lens for the wave functions
with frequency $\omega$ 
on the lens can be simply expressed as
\begin{equation}
\Psi_\mathrm{T}(\vec{x})=e^{-i\omega\frac{|\vec{x}|^2}{2f}}\Psi(\vec{x})\ ,
\label{tranwave}
\end{equation}
where $\vec{x}=(x,y)$ are coordinates on the lens located at $z=0$.
For example, an incident plane wave propagating along $z$-axis is written as
$\Psi(\vec{x}) = 1$ on the lens, whose phase does not depend on $\vec{x}$.
Then, from Eq.~(\ref{tranwave}), the transmitted wave becomes 
$\Psi_\mathrm{T}(\vec{x}) = 
\exp(- i\omega |\vec{x}|^2/2f)
\simeq 
\exp[-i\omega\sqrt{f^2+|\vec{x}|^2} + i\omega f]$.
This phase dependence describes that of the spherical wave converging on the
focus at $z=f$ in the Fresnel approximation ($|\vec{x}|\ll f$).

Let us consider a spherical screen located at $(x,y,z)=(x_\mathrm{S},y_\mathrm{S},z_\mathrm{S})$ with $x_\mathrm{S}^2+y_\mathrm{S}^2+z_\mathrm{S}^2=f^2$.
The transmitted wave converted by the lens is focusing and imaging on this screen.
When the transmitted wave on the lens is 
$\Psi_\mathrm{T}(\vec{x})$, 
the wave function on the screen is given by
\begin{equation}
 \Psi_\mathrm{S}(\vec{x}_\mathrm{S})=\int_{|\vec{x}|\leq d} d^2x \Psi_\mathrm{T}(\vec{x}) e^{i\omega L}\ .
\label{PsiI}
\end{equation}
where $d$ is the radius of the lens and 
$L$ is the distance between $(x,y,0)$ on the lens and 
$(x_\mathrm{S},y_\mathrm{S},z_\mathrm{S})$ on the screen: 
\begin{equation}
\begin{split}
 L&=\sqrt{(x_\mathrm{S}-x)^2+(y_\mathrm{S}-y)^2+z_\mathrm{S}^2}\\
&=\sqrt{f^2-2\vec{x}_\mathrm{S}\cdot\vec{x}+|\vec{x}|^2}
\simeq f-\frac{\vec{x}_\mathrm{S}\cdot\vec{x}}{f}+\frac{|\vec{x}|^2}{2f}\ ,
\end{split}
\label{dist}
\end{equation}
where $\vec{x}_\mathrm{S}=(x_\mathrm{S},y_\mathrm{S})$.
Substituting Eqs.~(\ref{tranwave}) and (\ref{dist}) into Eq.~(\ref{PsiI}), we obtain
\begin{equation}
 \Psi_\mathrm{S}(\vec{x}_\mathrm{S})\propto \int_{|\vec{x}|<d}d^2x\, \Psi(\vec{x})e^{-\frac{i\omega}{f}\vec{x}\cdot\vec{x}_\mathrm{S}}\ .
\label{lenstrans}
\end{equation}
This implies that the image on the screen can be obtained by the Fourier
transformation of the incident wave within a finite domain of the lens.
Equation~\eqref{lenstrans} motivates us to regard the observable defined in \eqref{lenstrans0} 
as the dual quantity of the Einstein ring.

\section{Null geodesics: geometrical optics}
\label{ngeo}
\begin{figure}
\begin{center}
\includegraphics[scale=0.7]{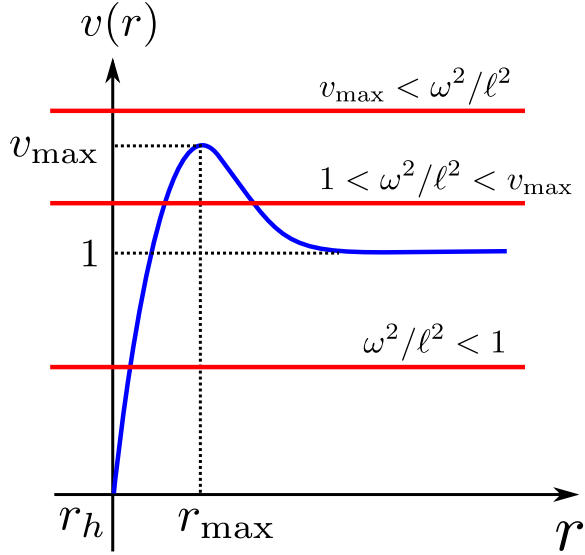}
\end{center}
\caption{
Typical profile of the effective potential.
For $1 < \omega^2/\ell^2 < v_\textrm{max}$, the orbit of null
 geodesics can go between two points on the AdS boundary ($r=\infty$). 
}
\label{pot_geo}
\end{figure}
To help intuitive understanding 
of the image of the AdS black hole which 
will be shown in the following sections, 
we consider null geodesics in the Sch-AdS$_4$ 
(that is, geometrical optics) 
in this section. 
In Appendix \ref{Val}, we derive the geodesic equation from the Klein-Gordon equation
and examine the validity of the Eikonal approximation in asymptotically AdS spacetimes.

\begin{figure}
  \centering
  \subfigure[$N_\textrm{w}=0$]
 {\includegraphics[scale=0.6]{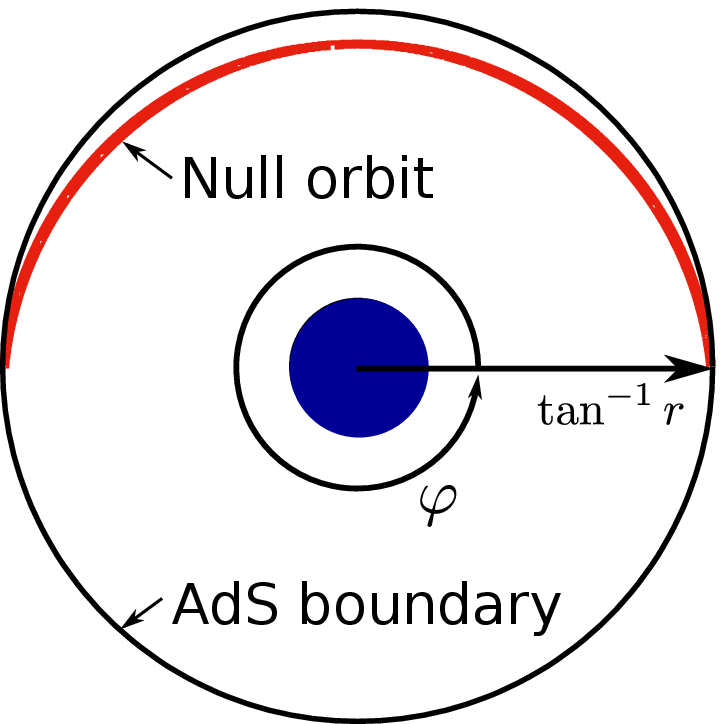}\label{orbit0}
  }
\subfigure[$N_\textrm{w}=1$]
 {\includegraphics[scale=0.6]{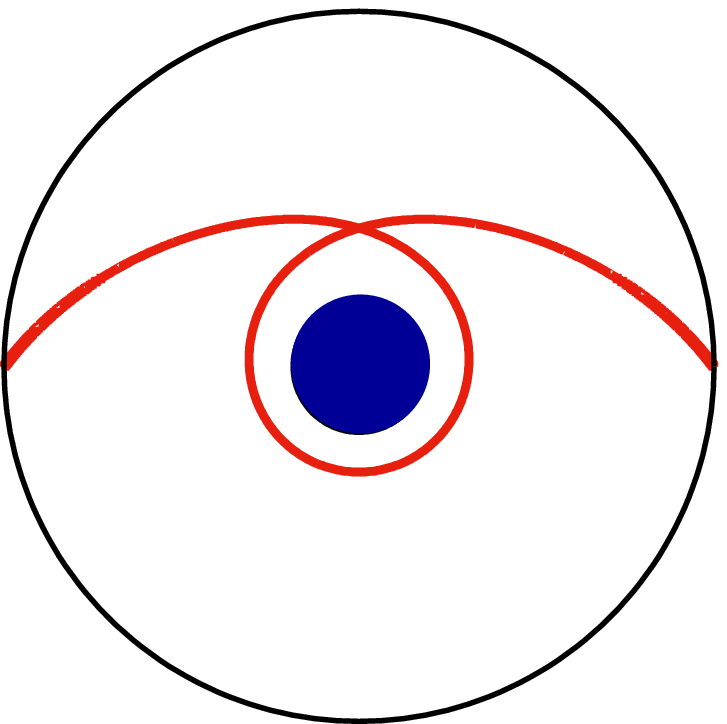}\label{orbit1}
  }
\subfigure[$N_\textrm{w}=2$]
 {\includegraphics[scale=0.6]{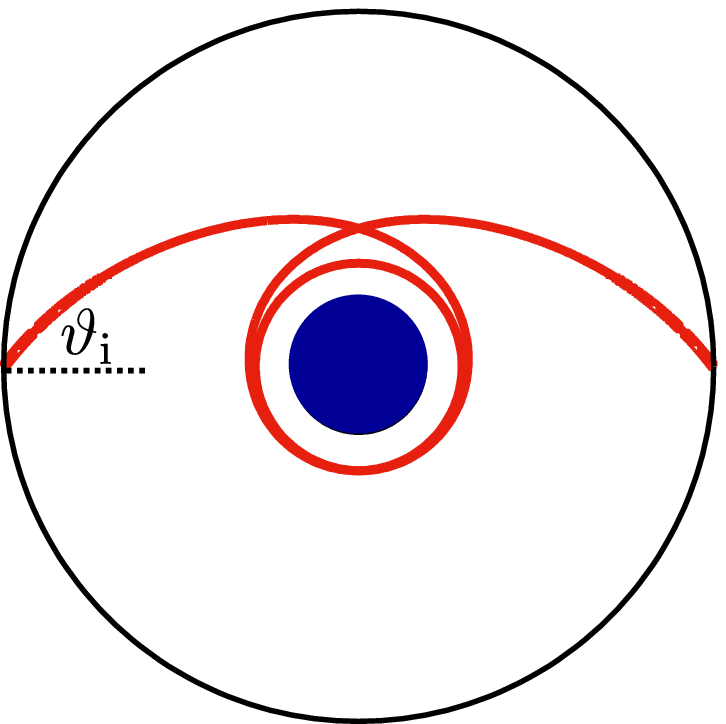}\label{orbit2}
  }
 \caption{
Null geodesics between $\varphi=0$ and $\pi$ for winding number $N_\textrm{w}=0,1,2$.
The horizon radius of the Sch-AdS$_4$ is fixed as $r_\mathrm{h}=0.3$.
$\vartheta_\mathrm{i}$ denotes the angle of incidence to the AdS boundary.
 }
 \label{geodesic_fig}
\end{figure}

It is well known that in the spherically symmetric spacetime an orbit of
geodesics lies in a plane passing through the center of the black hole.
Therefore, we can always rotate an orbital plane of the null geodesic to
coincide with the equatorial plane, without loss of generality.
In this section, for simplicity, we will study null geodesics on the
equatorial plane ($\theta=\pi/2$).
Then, the conserved energy and angular momentum are written as
\begin{equation}
 \omega=F(r)\dot{t}\ ,\quad 
 \ell=r^2\dot{\varphi}\ ,
\label{omegal}
\end{equation}
where ${}^\cdot=d/d\lambda$ and $\lambda$ is an affine parameter. From the null condition, we have
\begin{equation}
 \dot{r}^2=\omega^2-\ell^2 v(r)\ ,\quad v(r)\equiv \frac{F(r)}{r^2}\ .
\label{dotrsq}
\end{equation}
We note that, since affine parameters of null geodesics can be
rescaled by a constant factor, the null orbits depend only on the ratio of
the conserved quantities, $\ell/\omega$.
The effective potential $v(r)$ has a maximum value:
\begin{equation}
 v_\textrm{max}= \frac{(3r_\mathrm{h}^2+4)(3r_\mathrm{h}^2+1)^2}{27r_\mathrm{h}^2(r_\mathrm{h}^2+1)^2}\ ,
\label{vmax}
\end{equation}
at $r=r_\textrm{max}\equiv 3r_\mathrm{h}(r_\mathrm{h}^2+1)/2$.
The maximum of the effective potential corresponds to the photon sphere,
i.e., the unstable circular orbit of null geodesics.
The schematic functional profile of the effective potential $v(r)$ is shown in Fig.~\ref{pot_geo}.
We are now interested in null orbits between two points on the AdS
boundary. 
An observer is located at one point and a light source is at
the other point.
Then, we have to tune the parameters so that $1 < \omega^2/\ell^2 < v_\textrm{max}$ is satisfied.
When $\omega^2/\ell^2$ is close to (but less than) $v_\textrm{max}$, 
the geodesic goes through the vicinity of the photon sphere and can wind around the black hole.
Consequently, there exist infinite geodesics labeled by the winding
number, $N_\textrm{w}$, which connect fixed two points on the AdS boundary, 
and in the vicinity of the photon sphere infinitely many orbits accumulate.
Figure~\ref{geodesic_fig} shows 
some geodesics stretched between antipodal points on the AdS boundary for $r_\mathrm{h}=0.3$. 
In particular, for the geodesic with $N_\textrm{w}=\infty$ (i.e. geodesic from the photon sphere), the angular momentum per unit energy becomes
\begin{equation}
 \left(\frac{\ell}{\omega}\right)_\textrm{photon sphere}=\frac{1}{\sqrt{v_\textrm{max}}}=\frac{r_\mathrm{h}(r_\mathrm{h}^2+1)}{(r_\mathrm{h}^2+1/3)\sqrt{r_\mathrm{h}^2+4/3}}\ .
\label{ellps}
\end{equation}

We can naturally define the angle of incidence 
of the null geodesic to the AdS boundary by 
$\cos\vartheta_\mathrm{i} \equiv g_{ij} u^i n^j/(|u||n|)|_{r=\infty}$, 
where $u^i$ is
the spatial component of the 4-velocity of the geodesic, $n^i$ is the
normal vector to the AdS boundary and $g_{ij}$ is the induced metric on
the $t=\text{const.}$ surface.
($|u|$ and $|n|$ are the norms of $u^i$ and $n^i$ with
respect to $g_{ij}$.)
Using Eqs.~(\ref{omegal}) and (\ref{dotrsq}), we can explicitly
calculate the angle of incidence as 
\begin{equation}
 \sin\vartheta_\textrm{i} = \frac{\ell}{\omega}\ .
\label{incangle}
\end{equation}
Combining Eqs.~(\ref{ellps}) and (\ref{incangle}), 
we can determine the angle of incidence of the null geodesic from the photon sphere as a function of the horizon radius $r_\mathrm{h}$.
In the geometrical optics, this angle $\vartheta_\mathrm{i}$ gives the angular
distance of the image of the incident ray from the zenith if an observer
on the AdS boundary looks up into the AdS bulk.
If two end points of the geodesic and the center of the black hole are
in alignment, the observer see a ring image with a radius corresponding
to the incident angle $\vartheta_\mathrm{i}$ because of axisymmetry.

\section{Imaging AdS black holes}
\label{ImAdSBH}
\begin{figure}
\begin{center}
\includegraphics[scale=1.2]{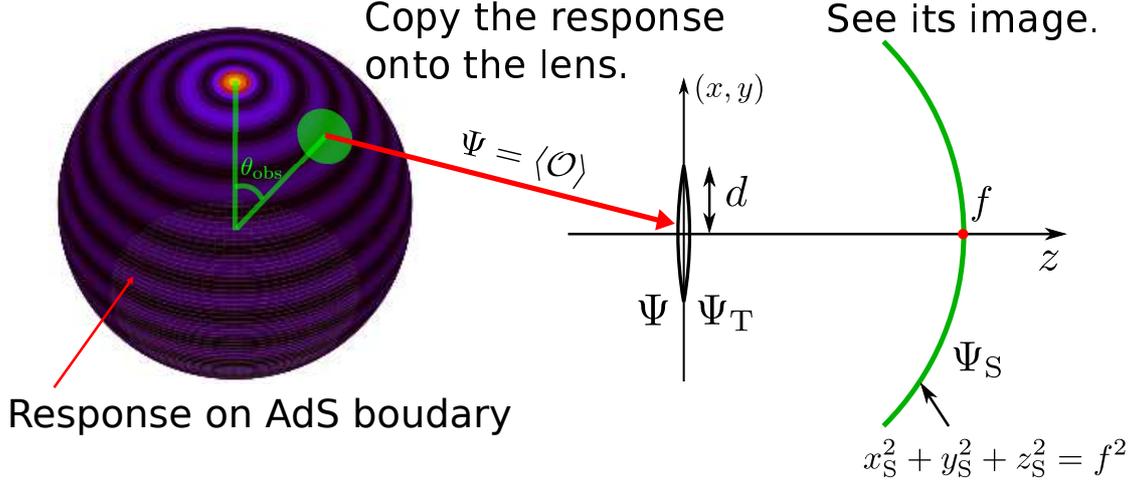}
\end{center}
\caption{
How to construct the image of the AdS black hole.
}
\label{restoim}
\end{figure}
Figure~\ref{restoim} shows our procedure to obtain the image of AdS black hole.
The sphere of the AdS boundary is depicted at the left side of the figure.
We show the absolute square of the response function $\langle \mathcal{O}(\theta)\rangle$ on the sphere
as the color map.
The brightest point is the north pole, i.e. the antipodal point of the Gaussian source.
The response function has the interference pattern caused by the diffraction of the wave by the black hole.
We now define an ``observation point'' at
$(\theta,\varphi)=(\theta_\textrm{obs},0)$ on the AdS boundary, where an
observer looks up into the AdS bulk.

To make an optical system, 
we virtually consider the flat 3-dimensional space $(x,y,z)$ 
as shown in the right side of Fig.~\ref{restoim}.
We set the convex lens on the $(x,y)$-plane.
The focal length and radius of the lens will be denoted by $f$ and $d$.
We also prepare the spherical screen at $(x,y,z)=(x_\mathrm{S},y_\mathrm{S},z_\mathrm{S})$ 
with $x_\mathrm{S}^2+y_\mathrm{S}^2+z_\mathrm{S}^2=f^2$. 
We copy the response function around the observation point 
as the wave function on the lens and observe its image.

We map the response function defined on $S^2$ onto the lens as follows.
We introduce new polar coordinates $(\theta',\varphi')$ as
\begin{equation}
 \sin\theta'\cos\varphi'+i\cos\theta'=e^{i\theta_\textrm{obs}}(\sin\theta\cos\varphi+i\cos\theta)\ ,
\end{equation}
such that the direction of the north pole is rotated to align with the
observation point: 
$\theta'=0 \Longleftrightarrow (\theta,\varphi)=(\theta_\textrm{obs},0)$.
Then, we define Cartesian coordinates as 
$(x,y)=(\theta'\cos\varphi',\theta'\sin \varphi')$
on the AdS boundary $S^2$.
In this coordinate system, we regard the response function around the observation point 
as the wave function on the lens: $\Psi(\vec{x})=\langle \mathcal{O}(\theta)\rangle$. 
The image on the screen can be obtained by Eq.~(\ref{lenstrans0}):
We perform the Fourier transformation within a finite domain on the
lens,
that is, applying an appropriate window function.

\begin{figure}
  \centering
  \subfigure[$r_\mathrm{h}=0.6$, $\theta_\textrm{obs}=0^\circ$]
 {\includegraphics[scale=0.46]{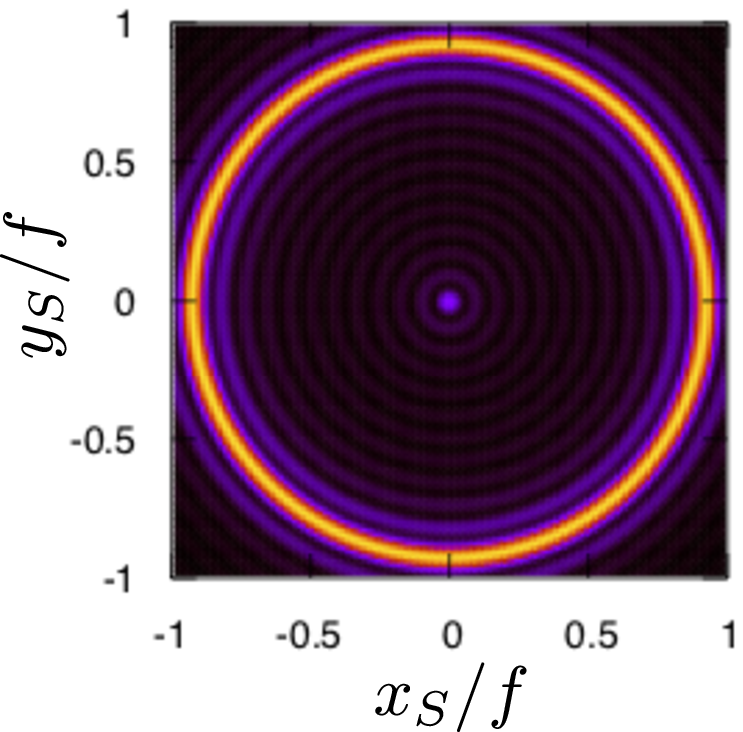}\label{rh06_angle0}
  }
\subfigure[$r_\mathrm{h}=0.6$, $\theta_\textrm{obs}\!=\!30^\circ$]
 {\includegraphics[scale=0.46]{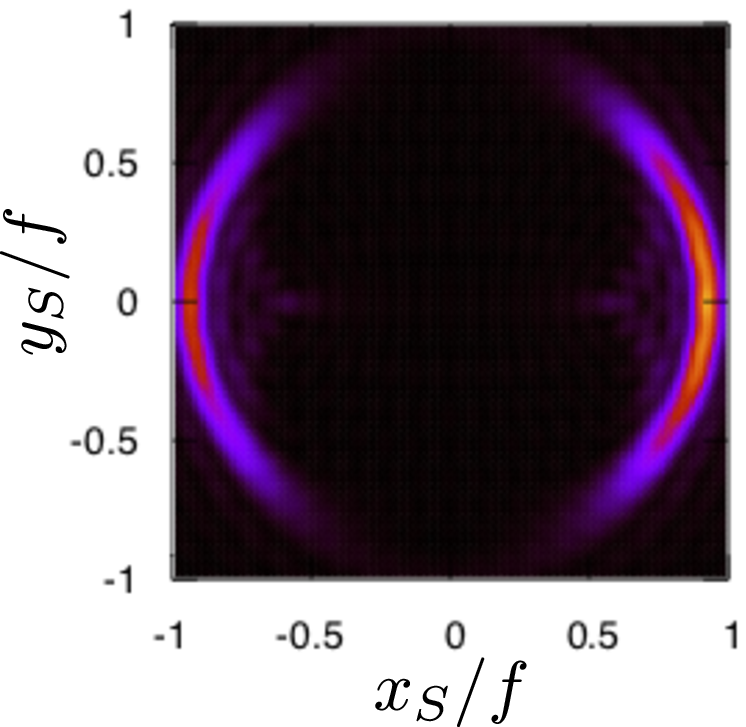}\label{rh06_angle30}
  }
\subfigure[$r_\mathrm{h}=0.6$, $\theta_\textrm{obs}=60^\circ$]
 {\includegraphics[scale=0.46]{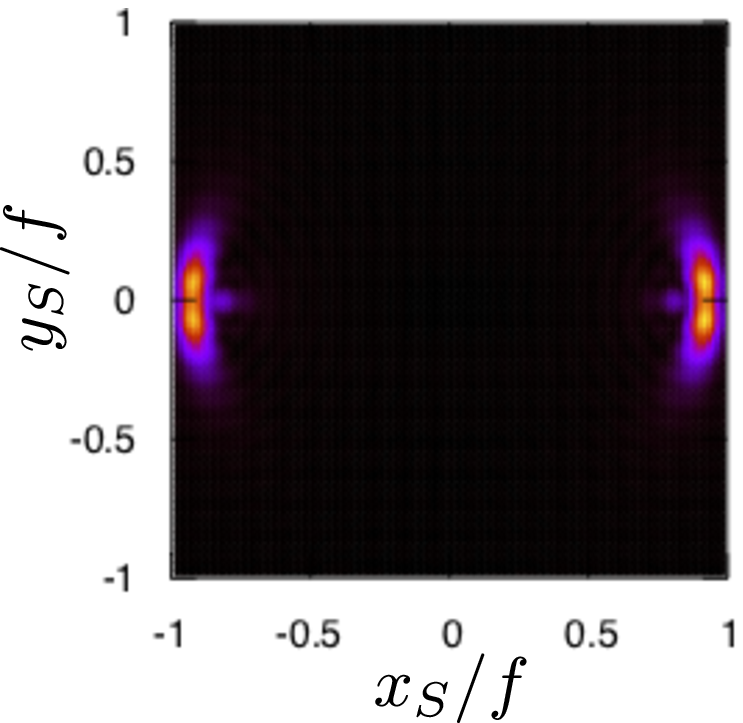}\label{rh06_angle60}
  }
\subfigure[$r_\mathrm{h}=0.6$, $\theta_\textrm{obs}=90^\circ$]
 {\includegraphics[scale=0.46]{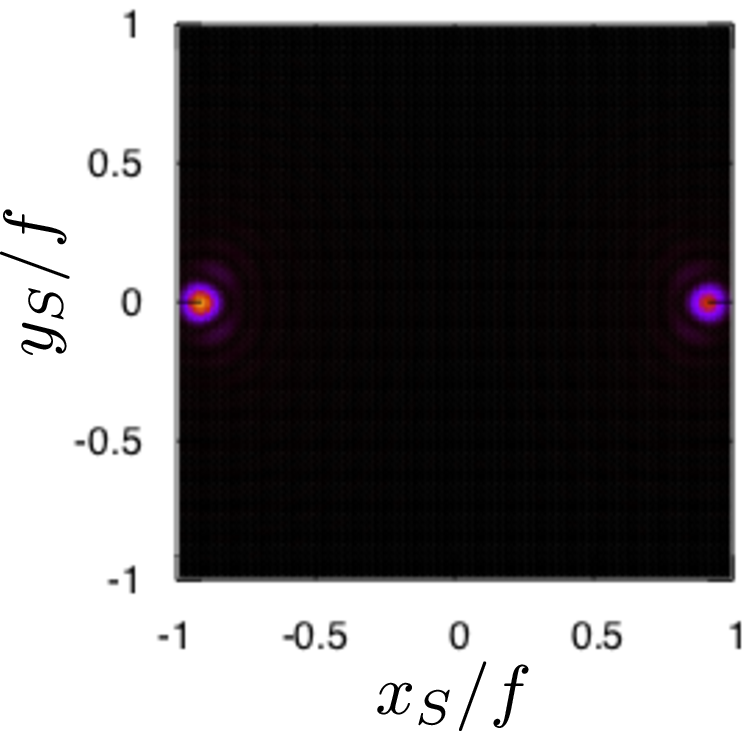}\label{rh06_angle90}
  }
\subfigure[$r_\mathrm{h}=0.3$, $\theta_\textrm{obs}=0^\circ$]
 {\includegraphics[scale=0.46]{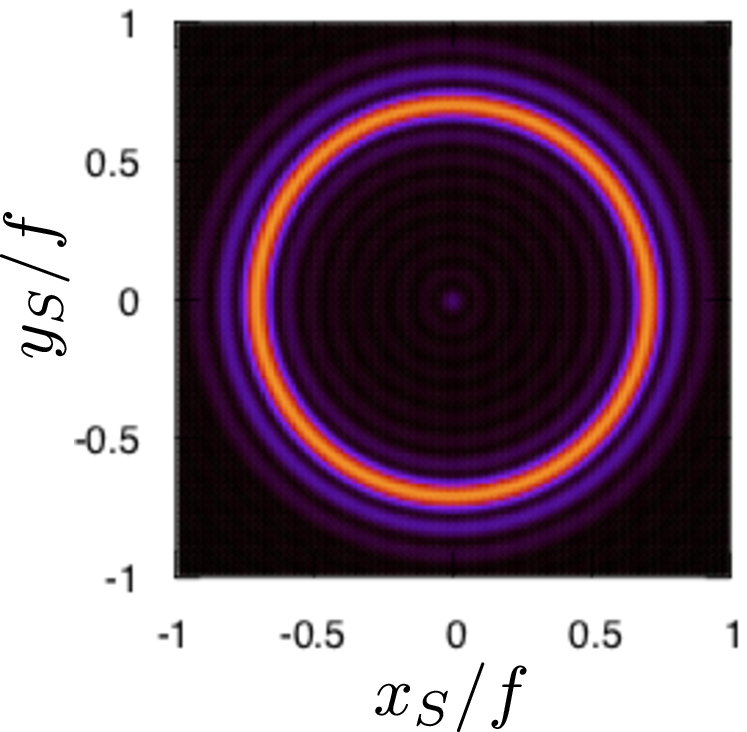}\label{rh03_angle0}
  }
\subfigure[$r_\mathrm{h}=0.3$, $\theta_\textrm{obs}=30^\circ$]
 {\includegraphics[scale=0.46]{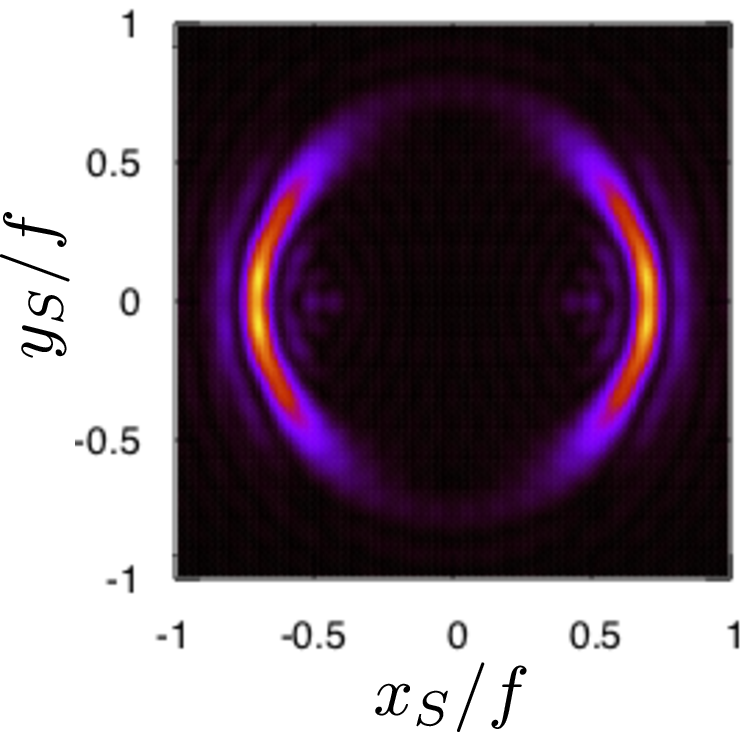}\label{rh03_angle30}
  }
\subfigure[$r_\mathrm{h}=0.3$, $\theta_\textrm{obs}=60^\circ$]
 {\includegraphics[scale=0.46]{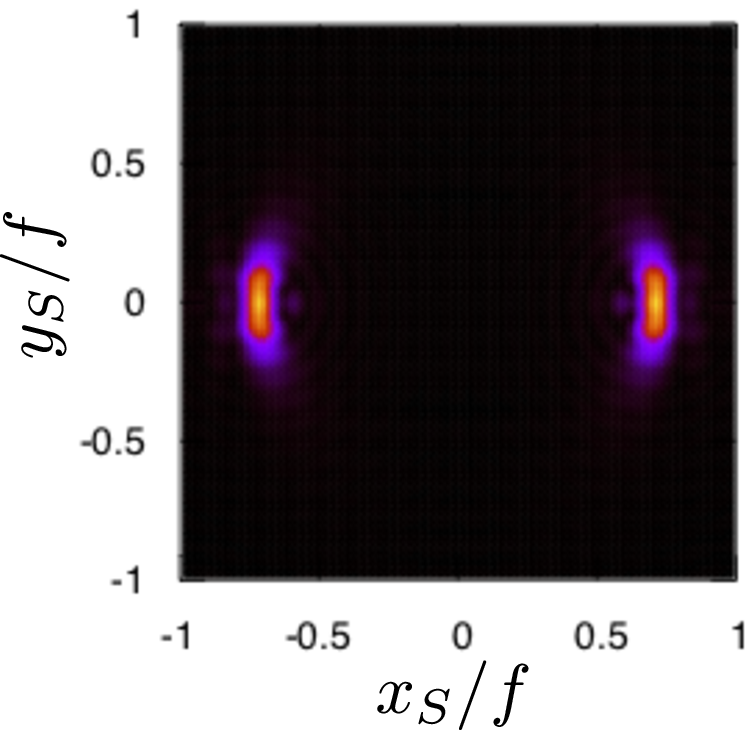}\label{rh03_angle60}
  }
\subfigure[$r_\mathrm{h}=0.3$, $\theta_\textrm{obs}=90^\circ$]
 {\includegraphics[scale=0.46]{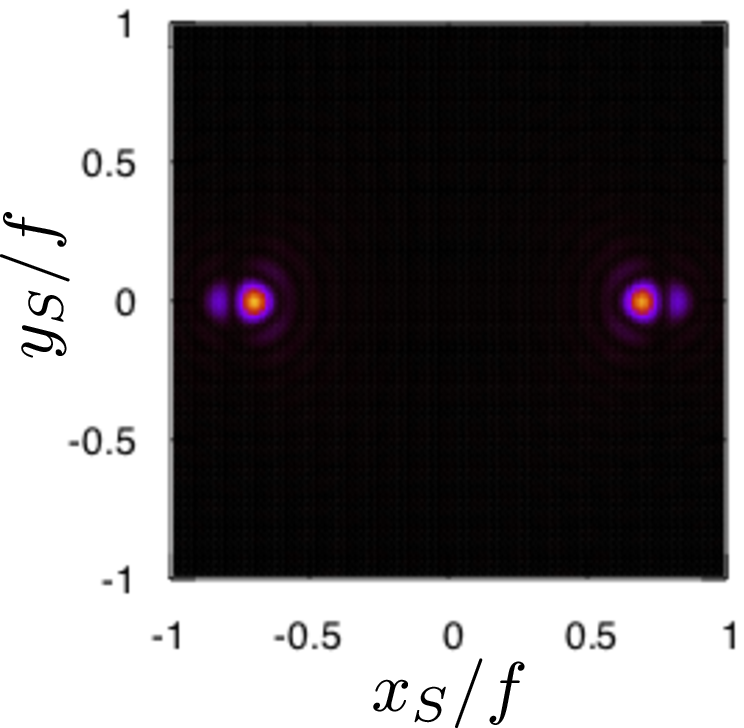}\label{rh03_angle90}
  }
\subfigure[$r_\mathrm{h}=0.1$, $\theta_\textrm{obs}=0^\circ$]
 {\includegraphics[scale=0.46]{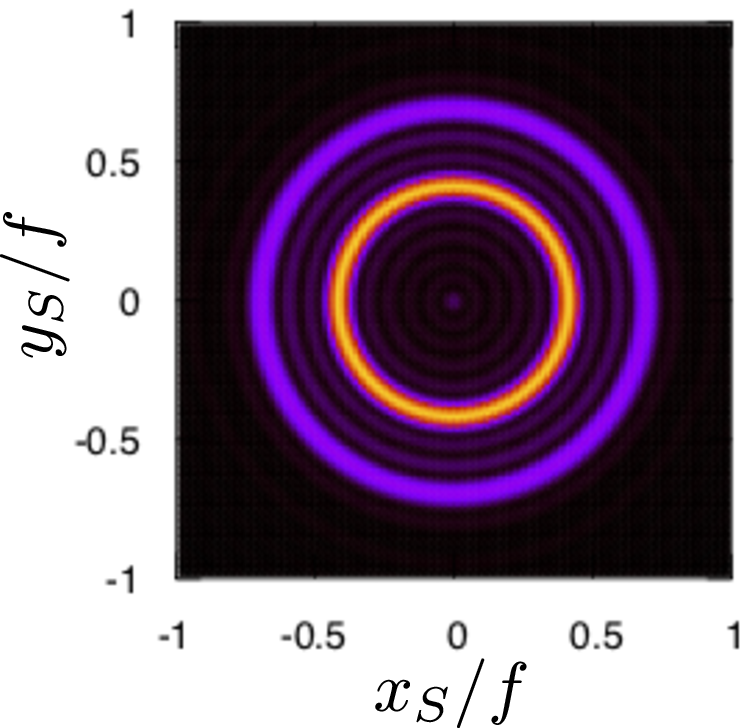}\label{rh01_angle0}
  }
\subfigure[$r_\mathrm{h}=0.1$, $\theta_\textrm{obs}=30^\circ$]
 {\includegraphics[scale=0.46]{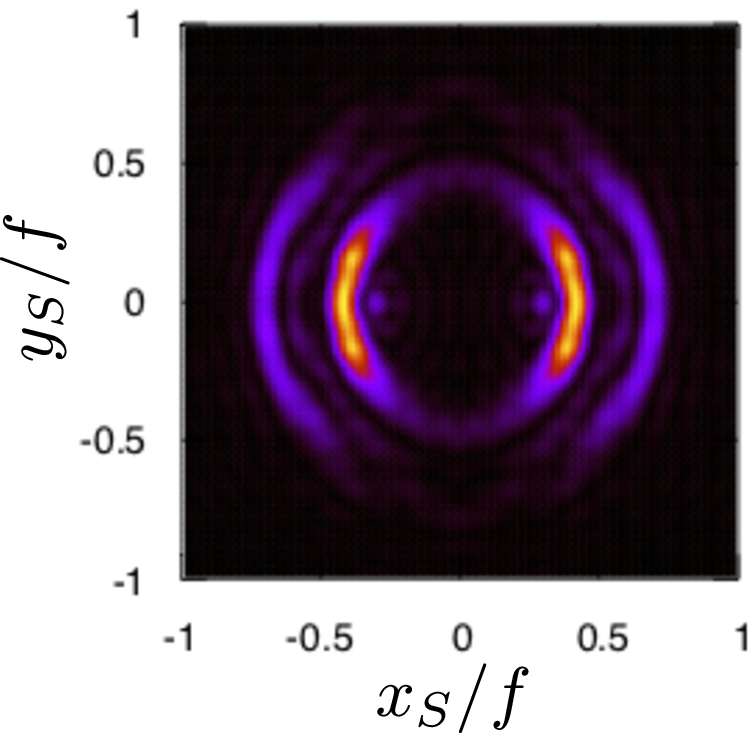}\label{rh01_angle30}
  }
\subfigure[$r_\mathrm{h}=0.1$, $\theta_\textrm{obs}=60^\circ$]
 {\includegraphics[scale=0.46]{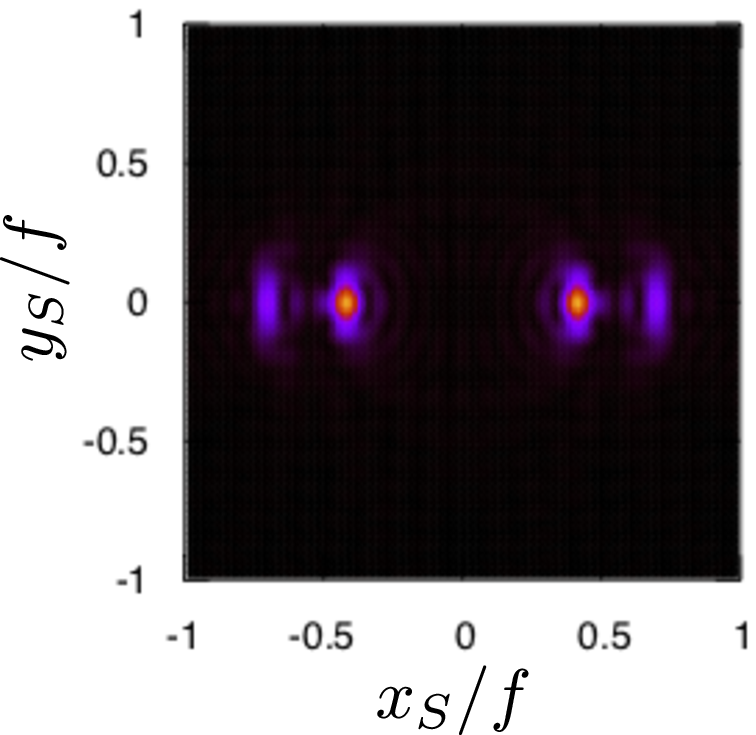}\label{rh01_angle60}
  }
\subfigure[$r_\mathrm{h}=0.1$, $\theta_\textrm{obs}=90^\circ$]
 {\includegraphics[scale=0.46]{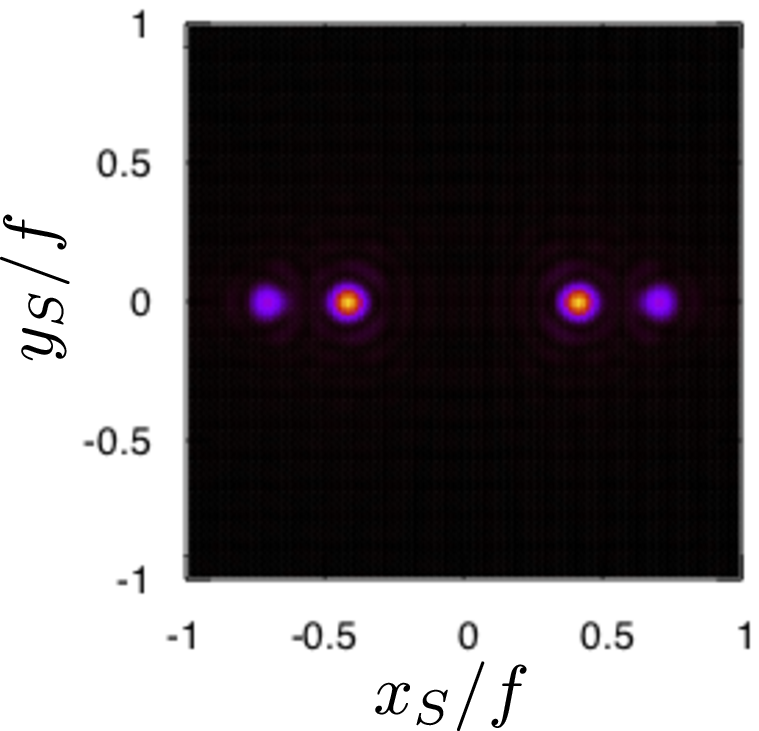}\label{rh01_angle90}
  }
 \caption{
Image of the Sch-AdS black hole for $\omega=80$, $\sigma=0.01$ and $d=0.5$.
The horizon radius and the observation point are varied as $r_\mathrm{h}=0.6,0.3,0.1$ and 
$\theta_\textrm{obs}=0^\circ,30^\circ,60^\circ,90^\circ$.
 }
 \label{rh010306_view}
\end{figure}

We now summarize our results on image formations of Sch-AdS black holes.
Figure~\ref{rh010306_view} shows images of the black hole observed at 
various observation points: 
$\theta_\textrm{obs}=0^\circ, 30^\circ, 60^\circ, 90^\circ$. 
The horizon radius is varied as $r_\mathrm{h}=0.6, 0.3, 0.1$.
We fix other parameters as $\omega=80$, $\sigma=0.01$ and $d=0.5$.
For $\theta_\textrm{obs}=0^\circ$, a clear ring is observed. 
As we will see for details in the next section, this ring corresponds to
the light rays from the vicinity of the photon sphere of the Sch-AdS$_4$. 
Not only the brightest ring, we can also see some concentric striped patterns. 
They are caused by a diffraction effect with imaging, which is not directly
related to properties of the black hole, because we find these patterns change
depending on the lens radius $d$ and frequency $\omega$.
As we change the angle of the observer, the ring is deformed. 
We observed similar images as those for asymptotically flat black hole~\cite{Nambu:2012wa,Kanai:2013rga,Nambu:2015aea}.
In particular, at $\theta_\textrm{obs}\sim 90^\circ$, we can observe double image of the point source.
They correspond to light rays which are clockwisely and anticlockwisely
winding around the black hole on the plane of $\varphi=0,\pi$.
The size of the ring becomes bigger as the horizon radius grows.

\begin{figure}
  \centering
  \subfigure[$\omega=10$]
 {\includegraphics[scale=0.6]{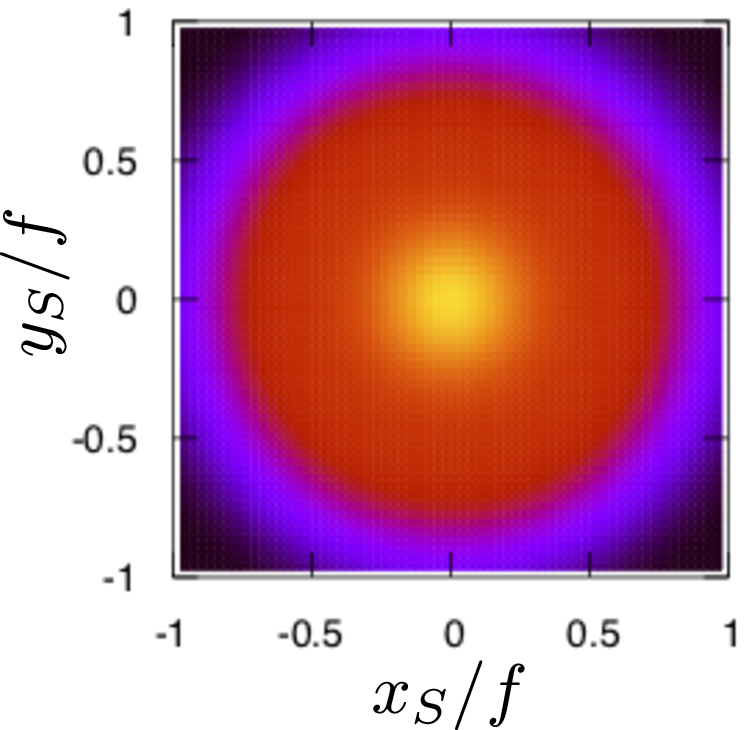}\label{rh03_om10}
  }
\subfigure[$\omega=20$]
 {\includegraphics[scale=0.6]{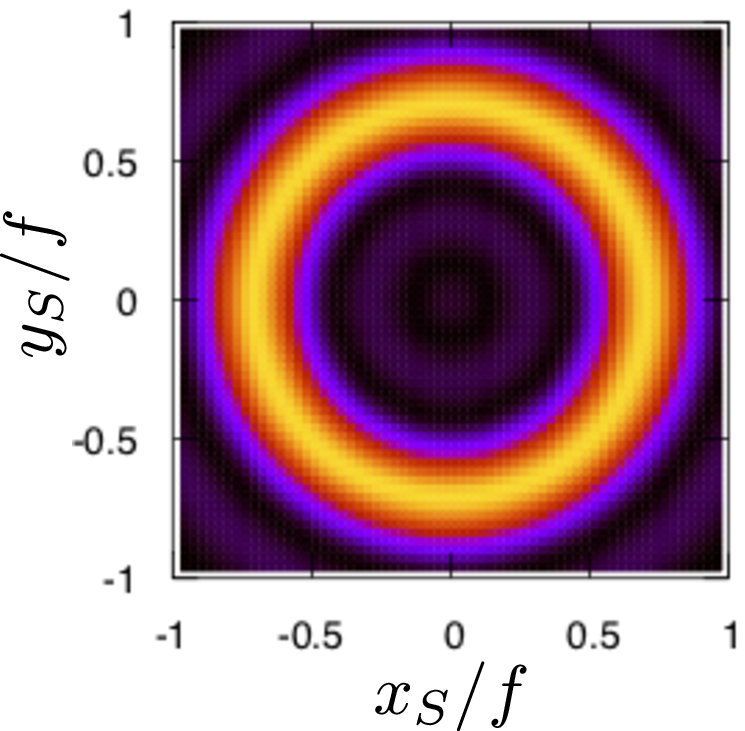}\label{rh03_om20}
  }
\subfigure[$\omega=30$]
 {\includegraphics[scale=0.6]{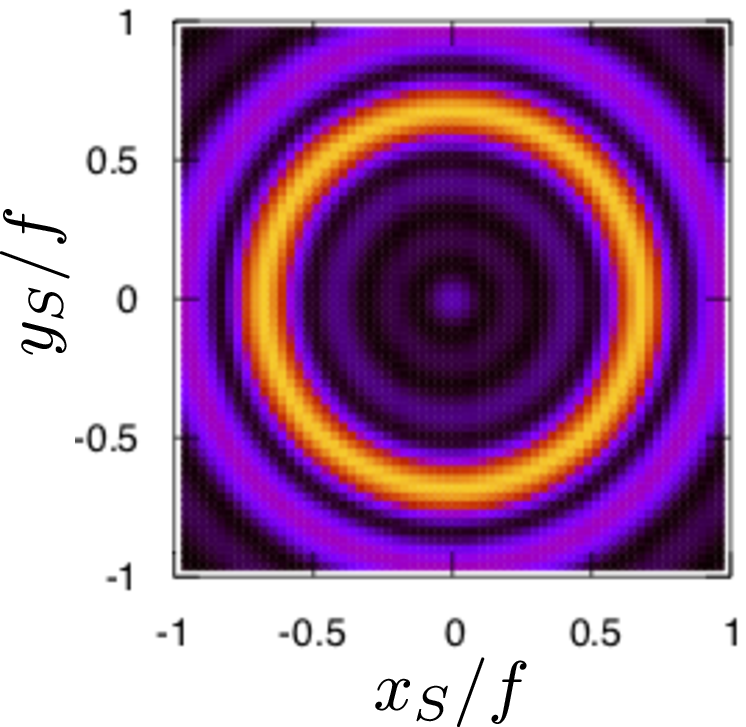}\label{rh03_om30}
  }
 \caption{
Frequency dependence of the image of the Sch-AdS black hole for $r_\mathrm{h}=0.3$, $\theta_\textrm{obs}=0$, $\sigma=0.01$ and $d=0.5$.
The frequency is varied as $\omega=10,20,30$.
 }
 \label{rh03_view_om}
\end{figure}

Figure~\ref{rh03_view_om} shows the 
the image of the Sch-AdS black hole for relatively small frequency, $\omega=10,20,30$.
Other parameters are fixed as $r_\mathrm{h}=0.3$, $\theta_\textrm{obs}=0$, $\sigma=0.01$ and $d=0.5$.
For $\omega=20,30$, 
we can only see the blurred ring because the wave effect is not negligible. 
For $\omega=10$, the ring disappears. 
In the geometric optics, we can only see the outside of the photon sphere.
In the wave optics, however, we have a chance to probe the region between photon sphere and event horizon due to some wave effects.
Do these images change if we modify the metric inside the photon sphere?
This is one of the interesting direction as a future work.

\section{Einstein radius}
\label{RingAngle}

\begin{figure}
\begin{center}
\includegraphics[scale=0.6]{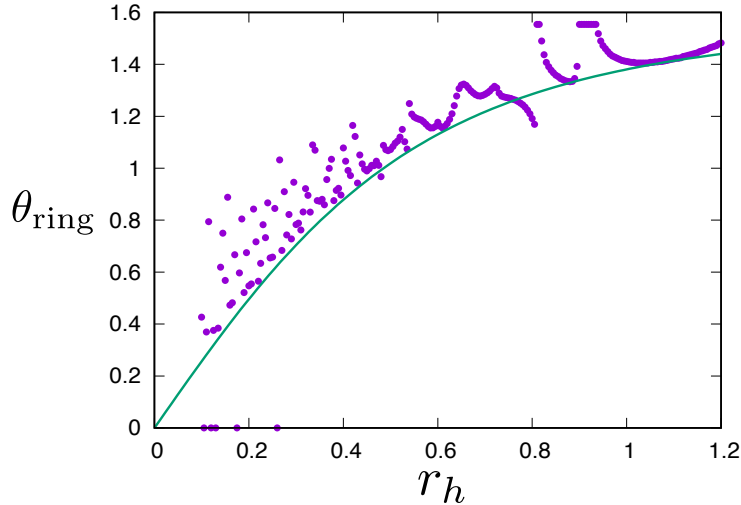}
\end{center}
\caption{
The Einstein radius $\theta_\textrm{ring}$ as a function of the horizon radius $r_\mathrm{h}$.
The green curve expresses the Einstein radius of the photon sphere, which is obtained by the geodesic approximation.
}
\label{thetaE}
 \end{figure}

To study the property of the brightest ring, 
we set the observation point at $\theta_\textrm{obs}=0$ and search 
$x_\mathrm{S}=x_\textrm{ring}$ at which $|\Psi_\mathrm{S}(\vec{x}_\mathrm{S})|^2$ has maximum value.
(Since the image has the rotational symmetry in $(x_\mathrm{S},y_\mathrm{S})$-plane for $\theta_\textrm{obs}=0$, 
we only focus on 
the $x_\mathrm{S}$-axis 
in this section.)
We will refer to
the angle of the Einstein ring  $\theta_\textrm{ring}=\sin^{-1}(x_\textrm{ring}/f)$ as the Einstein radius.
Figure~\ref{thetaE} shows the Einstein radius $\theta_\textrm{ring}$ by the purple points 
varying the horizon radius $r_\mathrm{h}$.
Note that the horizon radius relates to the total energy of the system by Eq.~(\ref{Mass}).
Although the Einstein radius fluctuates as the function of $r_\mathrm{h}$, 
it has an increasing trend as $r_\mathrm{h}$ is enlarged.

As we have seen in the geodesic analysis, 
there is an infinite number of
geodesics connecting antipodal points on the AdS boundary, which are
labeled by the winding number $N_\textrm{w}$.
Which geodesic in the geometrical optics corresponds to the ring found
in the image in the wave optics?
From the analysis of null geodesics, 
the (specific) angular momentum of the null
geodesic provides the angle of incidence $\vartheta_\mathrm{i}$ to the AdS boundary as shown in Eq.~(\ref{incangle}).
In the wave optics, the null geodesic with angular momentum $\ell$ should be 
a wave packet realized by the superposition of the spherical harmonics $Y_{\ell' 0}(\theta)$ with 
$\ell-\Delta \ell \leq \ell' \leq \ell+\Delta \ell$ ($\Delta \ell \ll \ell$).
For a sufficiently large $\ell$, 
the spherical harmonics behaves as $Y_{\ell 0}\sim e^{i\ell \theta}$.
Applying the Fourier transformation~(\ref{lenstrans}) onto the spherical harmonics, 
we have a peak in the image at $x_\mathrm{S}/f\simeq \ell/\omega$.
Thus, the angular distance of the image on the screen,
$\sin^{-1}(x_\mathrm{S}/f)$, coincides with the angle of incidence of
the null geodesic to the AdS boundary, 
$\vartheta_\mathrm{i} = \sin^{-1}(\ell/\omega)$.

Since we found that the angular momentum of the null geodesics from the
photon sphere is given by Eq.~(\ref{ellps}), they are expected to form the ring at 
$\theta_\textrm{ring}=\sin^{-1}(\ell/\omega)=\sin^{-1}(1/\sqrt{v_\textrm{max}})$.
The Einstein radius calculated from geodesic analysis 
is shown by the green curve in Fig.~\ref{thetaE}.
The curve seems to be consistent with the Einstein radius 
of the image constructed from the response function in the
wave optics.
This indicates that the major contribution to the brightest ring in the
image is originated by the ``light rays`` from the vicinity of the photon sphere, which
are infinitely accumulated.
Although there 
are expected to
be multiple Einstein rings corresponding to light lays with winding
numbers $N_\textrm{w}=0,1,2,\ldots$, the contribution for the image from small $N_\textrm{w}$ 
may be
so small that 
we cannot 
resolve 
them within our numerical accuracy.

The deviation of the Einstein radius $\theta_\textrm{ring}(r_\textrm{h})$ 
from the geodesic prediction can be considered as some wave effects.
In the AdS cases, whether the geometrical optics can adapt to imaging of
black holes is not so trivial even for a large value of $\omega$.
As studied in Appendix~\ref{Val}, 
the Eikonal approximation, which supports the geometrical optics, will
inevitably break down near the AdS boundary, while we have given
the source $J_\mathcal{O}$ and read the response 
$\langle\mathcal{O}\rangle$ on the AdS boundary.
Our results based on the wave optics imply that the geometrical
optics is qualitatively valid but gives a non-negligible deviation even
for a large $\omega$.

\section{Analytic examples of the images}
\label{Analy}

To illustrate our methods, in this section we provide examples in which 
images 
are obtained analytically. The first example is a $(2+1)$-dimensional CFT
on a sphere at zero temperature, which is dual to the pure AdS$_4$ geometry.
The second example is a $(1+1)$-dimensional CFT on a circle at a finite temperature,
which is dual to the BTZ black hole.

\subsection{Imaging AdS$_4$}
\label{ImAdS4}

In the absence of the black hole horizon, 
there should not be holographic images of the black hole.
To demonstrate it, here we study the case of pure AdS geometry. It is a gravity dual of
a CFT at zero temperature, and we have an analytic expression for the response function.

\begin{figure}
  \centering
  \subfigure[$\theta_\textrm{obs}=0^\circ$]
 {\includegraphics[scale=0.46]{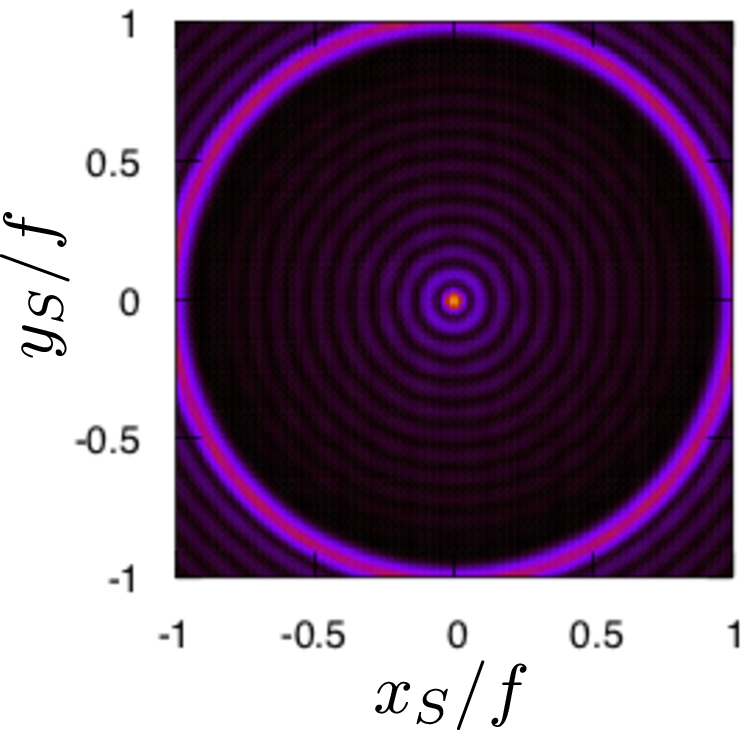}\label{rh00_angle0}
  }
\subfigure[$\theta_\textrm{obs}=30^\circ$]
 {\includegraphics[scale=0.46]{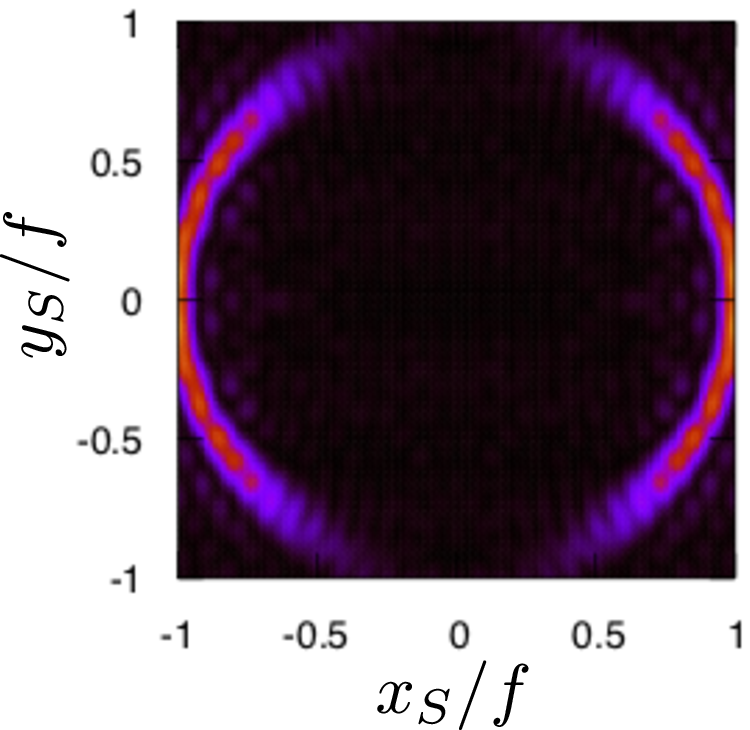}\label{rh00_angle30}
  }
\subfigure[$\theta_\textrm{obs}=60^\circ$]
 {\includegraphics[scale=0.46]{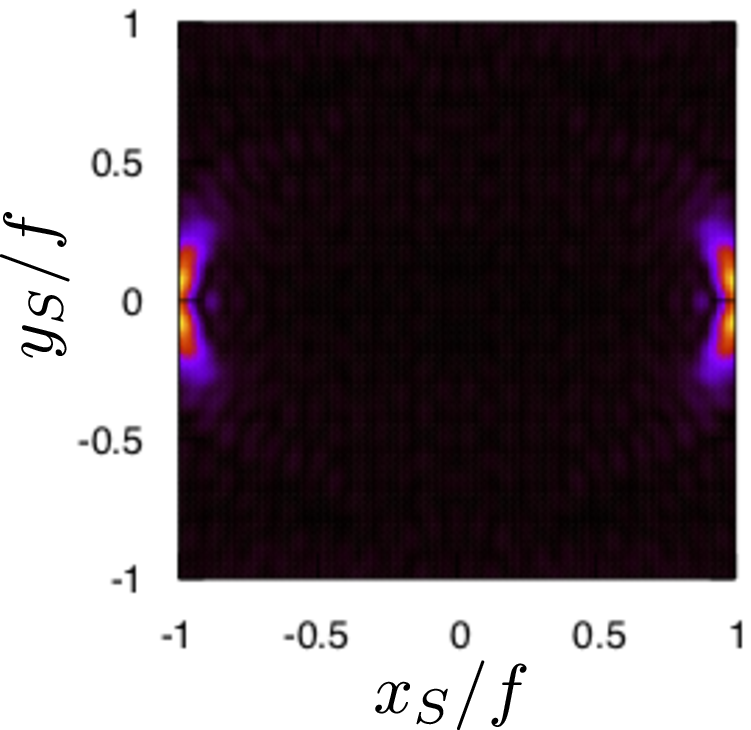}\label{rh00_angle60}
  }
\subfigure[$\theta_\textrm{obs}=90^\circ$]
 {\includegraphics[scale=0.46]{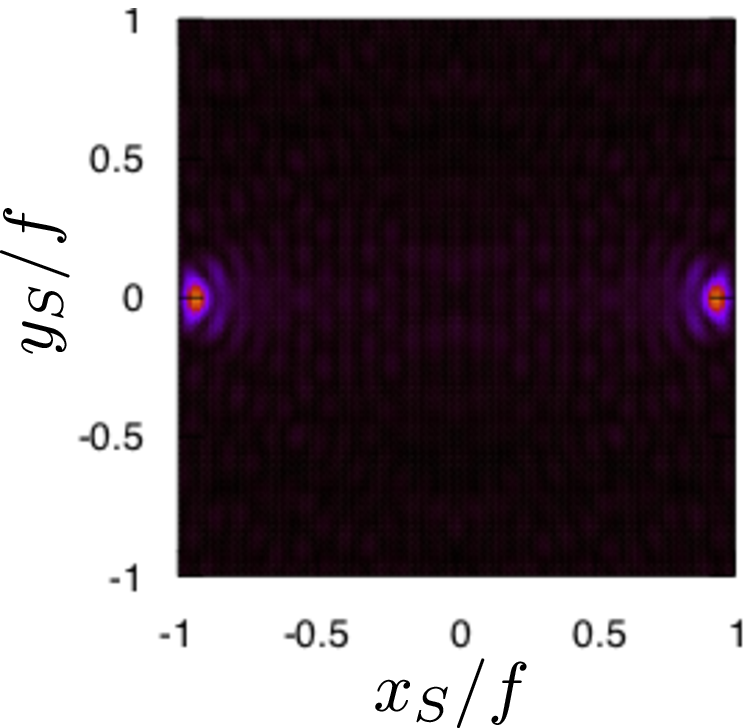}\label{rh00_angle90}
  }
 \caption{
Image of the pure AdS$_4$ ($r_\mathrm{h}=0$) for $\omega=80.5$, $\sigma=0.01$ and $d=0.5$.
The observation point are varied as 
$\theta_\textrm{obs}=0^\circ,30^\circ,60^\circ,90^\circ$.
 }
 \label{rh00_view}
\end{figure}

For the pure AdS$_4$ ($r_\mathrm{h}=0$), we can solve the scalar field equation analytically as
\begin{equation}
 \phi_\ell=\frac{2\Gamma(\frac{\ell+\omega+3}{2}) \Gamma(\frac{\ell-\omega+3}{2})}{\sqrt{\pi}\Gamma(\ell+\frac{3}{2})}
\left(\frac{r^2}{1+r^2}\right)^{\ell/2}F\left(\frac{\ell+\omega}{2},\frac{\ell-\omega}{2},\ell+\frac{3}{2};\frac{r^2}{1+r^2}\right)\ .
\end{equation}
Its asymptotic behavior is
\begin{equation}
 \phi_\ell=1+\frac{\omega^2-\ell(\ell+1)}{2r^2}+\frac{8\Gamma(\frac{\ell+\omega+3}{2}) \Gamma(\frac{\ell-\omega+3}{2})}
{3\Gamma(\frac{\ell+\omega}{2}) \Gamma(\frac{\ell-\omega}{2})}\frac{1}{r^3}+\cdots\ .
\end{equation}
Therefore, the response function is given by
\begin{equation}
 \langle \mathcal{O}(\theta)\rangle = \sum_\ell 
\frac{8\Gamma(\frac{\ell+\omega+3}{2}) \Gamma(\frac{\ell-\omega+3}{2})}
{3\Gamma(\frac{\ell+\omega}{2}) \Gamma(\frac{\ell-\omega}{2})}
c_\ell Y_{\ell 0}(\theta)\ ,
\label{pureAdSO}
\end{equation}
where 
$c_\ell=\int d\theta d\varphi \sin\theta\, g(\theta)Y_{l\,0}(\theta)$ whose explicit expression is in Eq.~(\ref{gexpand}).
The response diverges for $\omega=\ell+2n+3$ ($n=0,1,2,\cdots$).
This corresponds to the normal modes of the pure AdS$_4$. 
Applying the Fourier transformation~(\ref{lenstrans0}) onto Eq.~(\ref{pureAdSO}), 
we obtain the image of the pure AdS$_4$ as in Fig.~\ref{rh00_view}.
For $\theta_\textrm{obs}=0^\circ$, 
we can observe the bright spot at the center. 
This corresponds to the ``straight'' null geodesic from the south pole $\theta=\pi$ to the north pole $\theta=0$. This indicates that there is no black hole shadow in pure AdS$_4$ as expected.

Note that there exists
a ring in addition to the bright center. 
The angle seems $\theta_\textrm{ring}=\pi/2$.
What is the origin of the ring in the pure AdS?
For $r_\mathrm{h}=0$, the effective potential for null geodesics is simply given by
$v(r)=1+1/r^2$. This effective potential has the ``maximum'' at $r=\infty$.
Therefore, if we tune the angular momentum per unit energy as $\ell/\omega=1$, 
we can realize the null geodesic propagating along the AdS boundary ($r=\infty$).
As shown in \eqref{incangle} in Section \ref{RingAngle}, 
the incident angle $\vartheta_\mathrm{i} = \pi/2$ given by
$\ell/\omega=1$ 
corresponds to $\theta_\textrm{ring}=\pi/2$.
The ring found in Fig.~\ref{rh00_view} would be originate from the null lay along the AdS boundary.
Even for the Sch-AdS$_4$ spacetime, 
there should be null geodesics along the AdS boundary.
However, the ring formed by such geodesics is much weaker than the ring by the photon sphere.

\subsection{Imaging BTZ black holes}
\label{ImBTZ}

Another analytic example of the image is the $(1+1)$-dimensional CFT at a finite temperature.
As its gravity dual, we consider the BTZ black hole:
\begin{equation}
 ds^2=-(r^2-r_\mathrm{h}^2)dt^2+\frac{dr^2}{r^2-r_\mathrm{h}^2}+r^2d\varphi^2\ .
\label{BTZmetric}
\end{equation}
This spacetime is locally AdS and 
the Klein-Gordon equation $\Box\Phi=0$ can be analytically solved as~\cite{Cardoso:2001hn}
\begin{equation}
 \phi_m=\frac{\Gamma(a+1)\Gamma(b+1)}{\Gamma(a+b+1)}\, x^{-\frac{i\omega}{2r_\mathrm{h}}}\, F(a,b,1+a+b; x)\ ,
\label{BTZsol}
\end{equation}
where we decompose the scalar field as $\Phi=e^{-i\omega t + m\varphi}\phi_m(r)$ and define
\begin{equation}
 x=1-\frac{r^2}{r_\mathrm{h}^2}\ ,\quad
a=-\frac{i(\omega+m)}{2r_\mathrm{h}}\ ,\quad
b=-\frac{i(\omega-m)}{2r_\mathrm{h}}\ .
\end{equation}
This solution satisfies ingoing boundary condition at the horizon.
The asymptotic form of the solution near the AdS boundary is given by
\begin{equation}
\phi_m=1+\bigg[\frac{i\omega}{2r_\mathrm{h}}+ab\big\{\psi(a+1)+\psi(b+1)\big\}\bigg]\epsilon 
+ ab\, \epsilon \ln (e^{2\gamma-1}\epsilon) + \mathcal{O}(\epsilon^2\ln\epsilon)\ ,
\label{phiinfBTZ}
\end{equation}
where $\epsilon=1-x=r_\mathrm{h}^2/r^2$, $\gamma$ is the Euler's constant and $\psi(x)=\Gamma'(x)/\Gamma(x)$ is the polygamma function.
We consider the time-periodic Gaussian source around $\varphi=\pi$ as $\Phi|_{r=\infty}=e^{-i\omega t}g(\varphi)$ where $g$ is the Gaussian function introduced in Eq.~(\ref{BC}).
Then, the response function becomes 
\begin{equation}
 \langle \mathcal{O}(\varphi)\rangle = \sum_m \langle \mathcal{O}_m\rangle
c_m e^{im\varphi}\ ,\quad
\langle \mathcal{O}_m\rangle\equiv \left[\frac{i\omega}{2r_\mathrm{h}} + ab\big\{\psi(a+1)+\psi(b+1)\big\}\right]r_\mathrm{h}^2 \ ,
\end{equation}
where $c_m=(2\pi)^{-1/2}\int d\varphi g(\varphi) e^{im\varphi}\simeq (-1)^me^{-\sigma^2 m^2 /2}$.
Setting observation point at $\varphi=0$, we can obtain the image of the BTZ black hole as
\begin{equation}
 \Psi_\mathrm{S}(x_\mathrm{S})=\int_{-d}^d d\varphi \langle \mathcal{O}(\varphi)\rangle e^{-i\omega \varphi x_\mathrm{S}/f}
=2d \sum_m \langle \mathcal{O}_m\rangle
c_m \frac{\sin[d(m-\omega x_\mathrm{S}/f)]}{d(m-\omega x_\mathrm{S}/f)}\ .
\end{equation}
\begin{figure}
\begin{center}
\includegraphics[scale=0.5]{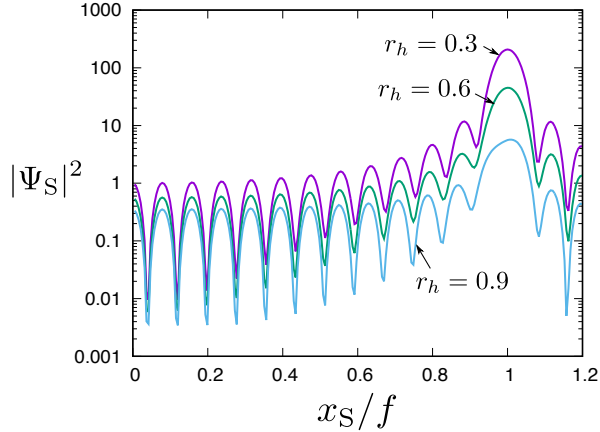}
\end{center}
\caption{
Image of BTZ black holes with $r_\mathrm{h}=0.3,0.6,0.9,1.2$.
}
\label{BTZimage}
\end{figure}
Figure~\ref{BTZimage} shows $|\Psi_\mathrm{S}|^2$ for $r_\mathrm{h}=0.3,0.6,0.9,1.2$.
We can always find peaks at $x_\mathrm{S}/f=1$ regardless of the horizon radius.
Corresponding angle is $\theta_\textrm{ring}=\pi/2$.
The effective potential for null geodesics in the BTZ spacetime is given by 
$v(r)=1-r_\mathrm{h}^2/r^2$.
This implies that all null geodesics with $\omega^2/\ell^2>1$ fall into the black hole.
However, like as the pure AdS case, 
if we tune the angular momentum per unit energy as $\ell/\omega=1$, 
we can realize the null geodesic propagating along the AdS boundary.
The ring found in the image of the BTZ would originate from the null geodesic on the boundary.
Therefore, the ``ring'' found in the image of BTZ does not have much information about the black hole spacetime. 
However, the dual gravitational theory does not need to be pure Einstein in general.
For example, 
it has been conjectured that the dual gravitational theory for a high $T_c$ superconductor is given by Einstein-Maxwell-charged scalar system~\cite{Gubser:2008px,Hartnoll:2008vx,Hartnoll:2008kx}.
Also, there can be a higher derivative corrections 
if we take into account the quantum gravity effect.
Then, the dual black hole space time will deviate from BTZ and 
we would be able to observe images of black holes even for $(1+1)$-dimensional materials.

\section{Einstein ring from retarded Green function}
\label{EringGreen}

We have been constructing the images of the AdS black hole
from the response functions in the previous sections.
Since the response function is closely related to the retarded Green
function, in this section we reinterpret the image of the Einstein ring
from holography in terms of the retarded Green function.
We demonstrate that poles of the Green function, which correspond to
  quasi-normal mode 
(QNM) 
frequencies 
on the AdS black hole 
in the gravity side, give major contribution for the formation of the Einstein ring.
  We also estimate the ``Einstein radius'' for weakly coupled quantum field theories which unlikely have their gravity duals.
  We find that the temperature dependence of the Einstein radius of weakly coupled theories is suppressed and gives qualitative difference with theories which have their gravitational duals.

The linear response $\langle \mathcal{O}\rangle$ with respect to the external source $J_\mathcal{O}$ on unit $S^2$
is written as 
\begin{equation}
 \langle \mathcal{O}(t,\theta)\rangle
= -2\pi\int dt' d\theta' \sin\theta' 
G(t,t',\theta,\theta')J_\mathcal{O}(t',\theta')\ ,
\label{eq:response_func}
\end{equation}
where we have assumed that the source $J_\mathcal{O}$ is axisymmetric. 
We introduce the retarded Green function $G(t,t',\theta,\theta')$.
It is well known that the Green function is given by the real time correlation function as
\begin{equation}
 G(t,t',\theta,\theta')=-i \Theta(t-t') \langle [\mathcal{O}(t,\theta),\mathcal{O}(t',\theta')]\rangle\ ,
\end{equation}
where $\Theta(t-t')$ is the step function and $\langle \cdots\rangle$ is the ensemble average with the equilibrium density matrix.
(For example, see Ref.~\cite{Natsuume:2014sfa} for the derivation.)
Let us suppose that the source $J_\mathcal{O}$ is monochromatic with
a frequency $\omega$.
The Green function and the source can be expanded in terms of 
Fourier modes 
and spherical harmonics 
$Y_\ell(\theta)\equiv Y_{\ell\,m=0}(\theta)$ as 
\begin{align}
 &G(t,t',\theta,\theta')=\sum_\ell \int \frac{d\omega'}{2\pi} e^{-i\omega' (t-t')} 
 G_\ell(\omega) Y_\ell(\theta)Y_\ell(\theta')\ ,\\
&J_\mathcal{O}(t',\theta')=e^{-i\omega t'}\sum_\ell J_\ell Y_\ell(\theta')\ .
\end{align}
Thus, we can rewrite the response function (\ref{eq:response_func}) as
\begin{equation}
 \langle \mathcal{O}(t,\theta)\rangle 
=-e^{-i\omega t} \sum_\ell  G_\ell(\omega) J_\ell Y_\ell(\theta)\ .
\label{OofG}
\end{equation}

For simplicity, we set the observation point at the north pole:
$\theta_\textrm{obs}=0$, which is the antipodal point of the  
external source localized at $\theta = \pi$.  
From the formula~(\ref{lenstrans0}) that we have proposed,
the image of the Einstein ring on the virtual screen is given by
\begin{equation}
 \Psi_S(t,\theta_\textrm{S})=\int_0^{2\pi} d\varphi \int_0^d d\theta\sin\theta
\langle \mathcal{O}(t,\theta)\rangle 
\exp\left(-\frac{i\omega}{f}\vec{x}\cdot \vec{x}_S\right)\ ,
\label{Psi_S}
\end{equation}
where $d$ is the radius of the lens and we assume $d\ll 1$ in the unit of the radius of $S^2$.
We have introduced polar coordinates on the boundary $S^2$ and the screen as
\begin{equation}
\begin{split}
&\vec{x}=\sin\theta ( \cos\varphi, \sin\varphi)\ ,\\
&\vec{x}_S=f\sin\theta_S (\cos\varphi_S,\sin\varphi_S)\ .
\end{split}
\end{equation}
Note that the formula (\ref{Psi_S}) means the Fourier transform of
the response function multiplied by a window function which is nonzero
only within a small finite region on $S^2$.

Now, we will rephrase the formula in terms of the retarded Green
function.
We perform the integration with respect to $\varphi$ by using 
$\vec{x}\cdot \vec{x}_S = f\sin \theta \sin\theta_S \cos(\varphi-\varphi_S)$,
and plug Eq.~(\ref{OofG}) into Eq.~(\ref{Psi_S}).
As a result, we obtain 
\begin{align}
 \Psi_S(t,\vec{x}_S)&= 
2\pi\int_0^d  d\theta \sin \theta \, \langle \mathcal{O}(t,\theta)\rangle \mathcal{J}_0(\omega \sin \theta_\textrm{S} \sin\theta)\nonumber\\
&= -2\pi e^{-i\omega t}  \sum_\ell  G_\ell(\omega) J_\ell \int_0^d  d\theta \sin \theta 
 Y_\ell(\theta) \mathcal{J}_0(\omega \sin \theta_\textrm{S} \sin\theta)\ ,
\label{PsiS2}
\end{align}
where $\mathcal{J}_n(x)$ is the Bessel function of the first kind, which comes from the $\varphi$-integration.
For $\theta \le d \ll 1$, the spherical harmonics can be approximated by the Bessel function as
\begin{equation}
 Y_\ell(\theta)\simeq \sqrt{\frac{\ell+1/2}{2\pi}}\mathcal{J}_0((\ell+1/2) \theta)\ .
\end{equation}
Using the above expression and replacing $\sin\theta\simeq \theta$ in Eq.~(\ref{PsiS2}), 
we can explicitly perform the $\theta$-integration and obtain
\begin{equation}
 \Psi_S(t,\vec{x}_S)
\simeq -\sqrt{2\pi}d^2 
e^{-i\omega t}  \sum_\ell  (\ell+1/2)^{1/2}G_\ell(\omega) \, J_\ell \, 
\Delta\left((\ell+1/2)d,\omega\sin\theta_\textrm{S} d\right)\ ,
\label{PsiS3}
\end{equation}
where we have defined 
\begin{equation}
 \Delta(x,y)\equiv \frac{x\mathcal{J}_1(x)\mathcal{J}_0(y)-y\mathcal{J}_1(y)\mathcal{J}_0(x)}{x^2-y^2}\ .
\end{equation}
The function $\Delta(x,y)$ has the highest peak with width $\sim \pi$ at
$x=y$ and is damping as $|x-y|$ becomes large.
If $\omega d \gg 1$, 
$\Delta\left((\ell+1/2)d,\omega\sin\theta_\textrm{S} d\right)$ in 
Eq.~(\ref{PsiS3}) has
greater values around 
$|\ell/\omega - \sin\theta_\textrm{S}| \lesssim \pi/\omega d \ll 1$.
Thus, we can eventually evaluate 
\begin{equation}
 \Psi_S(t,\vec{x}_S)
\propto
e^{-i\omega t} G_\ell(\omega) J_\ell\bigg|_{\ell=\omega\sin\theta_\textrm{S}}  \ .
\end{equation}
Formally, Eq.~(\ref{PsiS3}) means that the image on the virtual screen
corresponds to convolution of the Green function and the window function
in wave-number space $\ell$.
Therefore, it turns out that the image resolution is characterized by
the window function as 
$\Delta \ell/\omega \simeq \Delta \theta_\textrm{S} \simeq \pi/\omega d$.

In the view of the gravity side, poles of the retarded Green function in
the frequency domain correspond to 
QNM 
frequencies, $\omega=\Omega_\ell^n\in \bf{C}$, where $n=0,1,2,\ldots$ represent overtone numbers.
We can expect that $|\Psi_S|^2$ has a large value when the frequency
of the given monochromatic source, $\omega$, is close to the position of a
QNM frequency in the complex $\omega$-plane, so that the Einstein ring
is formed on the screen.
In other words, the condition
for the Einstein radius $\theta_\textrm{S}=\theta_\textrm{ring}$ is written as
\begin{equation}
 \omega\simeq \textrm{Re}\, \Omega^n_{\ell=\omega \sin\theta_\textrm{ring}} \ ,
  \label{thetaring_cond}
\end{equation}
for an overtone number $n$.

The field equation of the massless scalar field, 
$\Phi(t,r,\theta,\varphi) \equiv e^{-i\omega t}\sum_\ell Y_\ell(\theta) \psi_\ell(r)/r$,
is given by 
\begin{equation}
 \bigg[
 - \frac{d^2}{dr_*^2} + \ell(\ell+1)v(r) 
+ \frac{F(r)F'(r)}{r}
\bigg]\psi_\ell(r)
= \omega^2\psi_\ell(r) , 
\end{equation}
where $v(r)$ is the effective potential for null geodesics defined in Eq.~(\ref{dotrsq})
and 
we have introduced the tortoise coordinate $dr_* = dr/F(r)$.
According to the WKB analysis~\cite{Ferrari:1984zz},
QNMs are characterized by the behavior of the effective potential around
an extremum.
In the Eikonal limit $\omega \simeq \ell \gg 1$, the local maximum of
a part of the potential $v(r)$ plays a significant role 
and is given by Eq.~(\ref{vmax}).
As a result, the QNMs that originate from this local maximum 
are described by 
\begin{equation}
\begin{split}
&\textrm{Re}\, \Omega_\ell^n = \sqrt{\ell(\ell+1)v_\textrm{max}-\alpha^2}\ ,\\
&\textrm{Im}\, \Omega_\ell^n = \alpha\left(k(n)+\frac{1}{2}\right)\ ,
\end{split}
\end{equation}
where $\alpha\equiv \sqrt{-(d^2v/dr_\ast^2)/(2v)}|_{r=r_\textrm{max}}$
and $k(n)$ is a real number of $O(1)$.
Even though this expression of the QNM frequencies is derived for asymptotically flat spacetime, this is still valid for asymptotically AdS case 
since the QNM is highly oscillating as a function of $r_\ast$ for
$\omega ,\ell\gg 1$ and can be easily connect to the desired solution near the AdS boundary.
(For detailed WKB analysis in asymptotically AdS spacetimes, see Refs.~\cite{Festuccia:2008zx,Berti:2009wx,Berti:2009kk,Dias:2012tq}.)
For $\ell \gg 1$, we have $\textrm{Re}\, \Omega_\ell^n \simeq \ell\sqrt{v_\textrm{max}}$ and Eq.~(\ref{thetaring_cond}) gives
\begin{equation}
 \sin\theta_\textrm{ring}\simeq \frac{1}{\sqrt{v_\textrm{max}}}\ .
\end{equation}
This is consistent with our direct numerical calculations.

The retarded Green function $G_\ell(\omega)$ is a well-studied quantity in quantum field theories.
For example,  the Green function of a weakly coupled $\phi^4$ theory with mass $m$ and coupling $\lambda$
is given by
\begin{equation}
 G_\ell(\omega)=\frac{1}{-\omega^2+\ell(\ell+1)+m_T^2}\ ,
\end{equation}
where $m_T=m^2+\mathcal{O}(\lambda T^2)$ is the effective mass with the thermal effect. 
Then, the Einstein radius for weakly coupled theory is given by
$\sin^2\theta_\textrm{ring} = 1-m_T^2/\omega^2$. It does not depend on the temperature for a sufficiently large $\omega$ and gives $\theta_\textrm{ring}\simeq \pi/2$.
This suggests that, from the temperature dependence of the Einstein ring,
we can diagnose if a given quantum field theory has its gravity dual.

\section{Conclusion and Discussion}
\label{Concl}

We have studied how we can construct 
the holographic image of the AdS black hole using observables in its dual field theory.
We have considered a thermal CFT with a scalar operator, which corresponds in the AdS/CFT to
the massless scalar field in Sch-AdS$_4$ spacetime with a spherical horizon.
We  have put 
a time periodic localized source for the operator
and have
computed its response function by the AdS/CFT dictionary. 
Applying the Fourier transformation~(\ref{lenstrans0}) 
onto the response function, 
we have observed the Einstein ring as the image of the AdS black hole (Fig.~\ref{rh010306_view}).
The Einstein radius has an increasing trend as a function of the horizon radius $r_\mathrm{h}$.
It is also consistent with the angle of the photon sphere calculated from
the geodesic analysis on the basis of the geometrical optics.

We have shown 
that, if the dual black hole exists, 
we can construct 
the image of the AdS black hole from the observable in the thermal QFT.
In other words, being able to observe the image the AdS black hole in the thermal QFT
can be regarded as a necessary condition for the existence of the dual black hole.
Finding conditions for the existence of the dual gravity picture for a given quantum field theory is 
one of the most important problems 
in the AdS/CFT. 
We would be able to use the imaging of the AdS black hole as a test 
for it.
One of the possible  applications 
is  superconductors. 
It is known that some properties of high $T_c$ superconductors can be captured by the black hole physics
in AdS~\cite{Gubser:2008px,Hartnoll:2008vx,Hartnoll:2008kx}.
One of the 
other interesting applications 
is the Bose-Hubbard model.
It has been conjectured that the Bose-Hubbard model at the quantum critical regime has 
a gravity dual~\cite{Sachdev:2011wg,Fujita:2014mqa,BHMOTOC}. 
If we can realize these materials on $S^2$, 
they will be appropriate targets for observing AdS black holes by experiments.
Applying localized sources on such materials and measuring  their  
responses, 
we would be able to observe Einstein rings by tabletop experiments.

Can we distinguish the AdS black hole from thermal AdS by the observation of the Einstein ring?
Results in Section~\ref{ImAdS4} indicate that a ``ring'' will be also observed even for the thermal AdS.
The angle of the ring is, however, always fixed at $\theta_\textrm{ring}=\pi/2$
irrespective of 
the temperature of the thermal radiation in the global AdS. 
This implies that we would be able to distinguish the AdS black hole from thermal AdS 
by the temperature dependence of the Einstein radius.

Finally we make some comments on a relation to black hole chaos and AdS/CFT correspondence.
It has been suggested \cite{Maldacena:2015waa} that 
in quantum field theories with their gravity dual, the quantum Lyapunov
exponent $\lambda$ defined by out-of-time-ordered correlators saturates the bound $\lambda \leq 2\pi T$
where $T$ is the temperature of the system. The saturation is due to the strong red shift near the horizon 
of the black hole in the gravity dual \cite{Shenker:2013pqa}.
In our study, the photon sphere exists due to the strong curvature of the spacetime near the horizon,
and typical time delay is discretized because the winding number of the null geodesics characterizes
the observed image of the black hole. Since the time delay is nothing but the chaotic behavior of
the black hole, we expect some relation between the frequency $\omega$, temperature $T$
and the integration of the observed amplitude $\Psi_S$. It would be interesting to find a concrete
relation for these quantities, as another check for the existence of the gravity dual based on the choas, 
as well as the
direct imaging we proposed in this paper.

\vspace{5mm}
\begin{acknowledgments}
We would like to thank 
Paul Chesler, 
Vitor Cardoso, 
Sousuke Noda 
and
Chulmoon Yoo,
for useful discussions and comments.
We would also like to thank organizers and participants of the YITP workshop YITP-T-18-05 ``Dynamics in Strong Gravity Universe''
for the opportunity to present this work and useful comments.
The work of K.~H.~was supported 
in part by JSPS KAKENHI Grants No.~JP15H03658, 
No.~JP15K13483, and No.~JP17H06462. 
The work of K. M. was supported by JSPS KAKENHI Grant No. 15K17658 and 
in part by JSPS KAKENHI Grant No. JP17H06462.
The work of S.~K.~was supported in part by JSPS KAKENHI Grants No.~JP16K17704.
\end{acknowledgments}

\appendix

\section{Detail of numerical calculations}
\label{sec:detail}

The scalar field $\Phi(t,r,\theta,\varphi)$ is decomposed as 
\begin{equation}
 \Phi(t,r,\theta,\varphi)= e^{-i\omega t} \sum_{\ell=0}^\infty c_\ell\, \phi_\ell(r) Y_{\ell\,0}(\theta)\ ,
\label{Phidec}
\end{equation}
where $Y_{\ell\, 0}(\theta)$ is the scalar spherical harmonics with zero magnetic quantum number ($m=0$).
Since the boundary condition~(\ref{BC}) is axisymmetric, 
the scalar field can also be axisymmetric and decomposed only by $Y_{\ell\,0}(\theta)$.
Also the time dependence of the scalar field can be factorized by $e^{-i\omega t}$
because the boundary condition is monochromatic.
For actual computations, 
it is convenient to introduce the tortoise coordinate as
\begin{equation}
r_\ast = \int_\infty^r \frac{dr'}{F(r')}\ .
\end{equation}
In terms of the tortoise coordinate, 
$r_\ast=-\infty$ and $0$ correspond to the horizon $(r=r_\mathrm{h})$ and AdS boundary $(r=\infty)$, respectively.
Near the AdS boundary, the relation between $r_\ast$ and $r$ becomes 
\begin{equation}
 r_\ast=-\frac{1}{r}+\frac{1}{3r^3}+\mathcal{O}(\frac{1}{r^4})\ ,\quad
 r=-\frac{1}{r_\ast}+\frac{r_\ast}{3}+\mathcal{O}(r_\ast^2)\ .
\label{rast_r_asym}
\end{equation}
Substituting Eq.~(\ref{Phidec}) into Eq.~(\ref{scalareq0}), we obtain the equation for $\phi_\ell$ as
\begin{equation}
 \left[\frac{d^2}{dr_\ast^2}+\frac{2F}{r}\frac{d}{dr_\ast}  +\left(\omega^2-\frac{\ell(\ell+1)F}{r^2}\right)\right]\phi_\ell=0\ .
\label{scalareq2}
\end{equation}
The asymptotic solution at horizon becomes
\begin{equation}
 \phi_\ell\sim e^{-i\omega r_\ast}\quad (r\to r_\mathrm{h})\ ,
\label{ingoing}
\end{equation}
where we took the ingoing mode. 
The asymptotic solution at the AdS boundary is given by
\begin{equation}
\phi_\ell= p_0 (1+p_2 r_\ast^2+p_4r_\ast^4+p_5r_\ast^5+\cdots) + q_3 (r_\ast^3+q_5 r_\ast^5+\cdots)\ .
\label{phiinf}
\end{equation}
where
\begin{equation}
\begin{split}
&p_2=\frac{1}{2}\{\omega^2-\ell(\ell+1)\}\ ,\quad
p_4=\frac{1}{24}[-3\{\omega^2-\ell(\ell+1)\}^2+2\{4\omega^2-\ell(\ell+1)\}]\ ,\\
&p_5=\frac{1}{20}r_\mathrm{h}(1+r_\mathrm{h}^2)\{3\omega^2-\ell(\ell+1)\}\ ,\quad 
q_5=\frac{1}{10}\{-\omega^2+\ell(\ell+1)+4\}\ .
\end{split}
\end{equation}
Here, $p_0$ and $q_3$ are not determined by the asymptotic expansion, which correspond to the source and response.

We numerically integrate Eq.~(\ref{scalareq2}) from horizon to AdS boundary.  
From the ingoing condition~(\ref{ingoing}), 
we impose the initial condition at $r_\ast = r_\ast^\textrm{min} \simeq -3$ as
\begin{equation}
\phi_\ell=1\ ,\quad
\frac{d\phi_\ell}{dr_\ast}=-i\omega\ .
\end{equation}
Solving Eq.~(\ref{scalareq2}) by the 4th order Runge-Kutta method, we obtain 
numerical values of $\phi$ and $d\phi/dr_\ast$ near the AdS boundary, $r_\ast=r_\ast^\textrm{max}\simeq -10^{-3}$.
From Eq.~(\ref{phiinf}), we obtain the boundary value of the scalar field  as
\begin{equation}
 p_0=\left.\frac{\phi_\ell}{1+p_2 r_\ast^2}\right|_{r_\ast=r_\ast^\textrm{max}}
+\mathcal{O}((r_\ast^\textrm{max})^3)\ .
\end{equation}
Using the obtained complex asymptotic value $p_0$, we normalize the numerical solution as
\begin{equation}
 \bar{\phi}_\ell(r)=\frac{\phi_\ell(r)}{p_0}\ .
\end{equation}
This solution satisfies $\bar{\phi}_\ell\to 1$ ($r\to \infty$).
For notational simplicity, hereafter, 
we will omit the ``bar'' of the scalar field.

To obtain the response $q_3$, 
we differentiate Eq.~(\ref{scalareq2}) by $r_\ast$.
Then, $d^n\phi_\ell/r_\ast^n$ ($n=0,1,2,3$) appear in the resultant equation.
The second derivative $d^2\phi_\ell/r_\ast^2$ can be eliminated by Eq.~(\ref{scalareq2}).
As the result, 
we can compute $d^3\phi_\ell/r_\ast^3$ from our numerical data of $d\phi_\ell/r_\ast$ and $\phi_\ell$.
From the third order derivative, we obtain $q_3$ as
\begin{equation}
 q_3=\left.\frac{d^3\phi_\ell/r_\ast^3-24 p_4 r_\ast - 60 p_5 r_\ast^2}{6\{1+10 q_5 r_\ast^2\}}\right|_{r_\ast=r_\ast^\textrm{max}}
+\mathcal{O}((r_\ast^\textrm{max})^3)\ .
\end{equation}

For $\sigma\ll 1$, we can decompose the Gaussian source by the spherical harmonics as
\begin{equation}
 g(\theta)=\sum_\ell c_\ell \, Y_{\ell 0}(\theta)\ ,\quad 
c_\ell\simeq (-1)^\ell\sqrt{\frac{\ell+1/2}{2\pi}}\exp\left(-\frac{1}{2}(\ell+1/2)^2\sigma^2\right)\ .
\label{gexpand}
\end{equation}
Therefore, the response function in $\theta$-space is given by
\begin{equation}
 \langle \mathcal{O}(\theta)\rangle =-\sum_\ell c_\ell\, q_3\, Y_{\ell 0}(\theta)\ .
\end{equation}
Substituting this into Eq.~(\ref{lenstrans0}) and performing the 2-dimensional Fourier transformation, 
we obtain the images in Fig.~\ref{rh010306_view}.

\section{Validity of geometrical optics approximation in AdS}
\label{Val}

It is well known that, if we adopt the Eikonal approximation, the
massless Klein-Gordon equation yields the Hamilton-Jacobi equation for
null geodesic in general curved spacetime. 

We assume that the scalar field is $\Phi = a(x^\mu) e^{iS(x^\mu)}$ and
the gradient of the phase function, $\partial_\mu S$, has typically 
the same scale as frequency $\omega$.
Substituting this ansatz into the field equation and taking 
$\omega \gg 1$, we have the Eikonal equation from the leading in
$\omega$ as 
\begin{equation}
 g^{\mu\nu}\partial_\mu S \partial_\nu S = 0 .
  \label{eq:Eikonal}
\end{equation}
This is nothing but the Hamilton-Jacobi equation for
massless particle, where $p_\mu \equiv \partial_\mu S$ is the $4$-momentum.
In fact, differentiating Eq.~(\ref{eq:Eikonal}), 
we can easily derive null geodesic equation as 
\begin{equation}
 0 = \nabla_\alpha (g^{\mu\nu}\partial_\mu S \partial_\nu S)
  = 2 p^\mu \nabla_\mu p_\alpha 
\end{equation}
together with the null condition $g^{\mu\nu}p_\mu p_\nu = 0$.

For the null geodesic equation, 
we consider the solution with the conserved energy and angular
momentum given by (\ref{omegal}), 
so that Hamilton's function is 
\begin{equation}
 S(t,r,\theta,\varphi) = - \omega t + \ell\varphi
  + \int\frac{dr}{F(r)}\sqrt{\omega^2 - \ell^2 v(r)} ,
\end{equation}
where 
\begin{equation}
 \frac{\partial S}{\partial t} = p_t = -\omega ,\quad
\frac{\partial S}{\partial \varphi} = p_\varphi = \ell ,\quad 
\frac{\partial S}{\partial r} = p_r = 
\frac{1}{F(r)}\sqrt{\omega^2 - \ell^2 v(r)} .
\label{eq:conjugate_momentum}
\end{equation}
Thus, it turns out that the wave front characterized by $S(x^\mu)$ will
propagate along trajectories of null geodesics under the Eikonal
approximation, that is, the geometrical optics.

Now, let us deal with the field equation more continously.
We focus on an eigenstate with frequency $\omega$ and angular momentum $\ell$.
By defining $\psi \equiv r\phi_\ell$, 
the field equation (\ref{scalareq2}) can be rewritten  
as  
\begin{equation}
 \left[\frac{d^2}{dr_\ast^2}-V(r)\right]\psi = 0\ ,
\label{scalareq3}
\end{equation}
where the effective potential is 
\begin{equation}
 V(r)=-\omega^2+l(l+1)\frac{F(r)}{r^2}+\frac{F(r)F'(r)}{r}\ .
\end{equation}
If $|V(r)| \gg 1$, we obtain the WKB solution  
\begin{equation}
\psi = \frac{1}{\sqrt{p_{r_\ast}}}
 \exp\left(i\int^{r_\ast}\!\!dx \, p_{r_\ast}(x) \right) ,
\end{equation}
where 
\begin{equation}
 p_{r_\ast}^2 = - V(r)
\end{equation}
Compared with (\ref{eq:conjugate_momentum}), we find the following
correspondence: 
\begin{equation}
 p_{r_\ast} \Longleftrightarrow 
  \frac{\partial S}{\partial r_\ast} = F(r)\frac{\partial S}{\partial r} .
\end{equation}
If the condition 
\begin{equation}
 \omega^2, \ell^2 \gg \frac{F(r)F'(r)}{r} 
\end{equation}
is satisfied, 
both the Eikonal approximation and the WKB approximation lead to the
same effective potential and phase function for $\omega, \ell \gg 1$.
However, since $F(r)F'(r)/r \simeq 2r^2$ near the AdS boundary, even for
any large frequency $\omega$ the above condition will violate as goes to
the AdS boundary.
Roughly speaking, this is because the gradient of the metric function, 
$\partial_r g_{\mu\nu}$, will be larger than the frequency due to a
so-called AdS potential. 

As a result, in order to predict the wave propagation of the
scalar field, inside the AdS bulk including the black hole we can apply
the geometric optics for a sufficiently large frequency $\omega$.
Near the AdS boundary, we should deal with it in the WKB method
together with appropriate matching procedures.


\begin{thebibliography}{99}

\bibitem{Maldacena:1997re}
  J.~M.~Maldacena,
  ``The Large N limit of superconformal field theories and supergravity,''
  Int.\ J.\ Theor.\ Phys.\  {\bf 38} (1999) 1113
   [Adv.\ Theor.\ Math.\ Phys.\  {\bf 2} (1998) 231]
  [hep-th/9711200].

\bibitem{Gubser:1998bc}
  S.~S.~Gubser, I.~R.~Klebanov and A.~M.~Polyakov,
  ``Gauge theory correlators from noncritical string theory,''
  Phys.\ Lett.\ B {\bf 428} (1998) 105
  [hep-th/9802109].

\bibitem{Witten:1998qj}
  E.~Witten,
  ``Anti-de Sitter space and holography,''
  Adv.\ Theor.\ Math.\ Phys.\  {\bf 2} (1998) 253
  [hep-th/9802150].



\bibitem{Falcke:1999pj}
  H.~Falcke, F.~Melia and E.~Agol,
  ``Viewing the shadow of the black hole at the galactic center,''
  Astrophys.\ J.\  {\bf 528} (2000) L13
  [astro-ph/9912263].

\bibitem{EHT}
  K.~Akiyama {\it et al.} [Event Horizon Telescope Collaboration],
  ``First M87 Event Horizon Telescope Results. I. The Shadow of the Supermassive Black Hole,''
  Astrophys.\ J.\  {\bf 875}, no. 1, L1 (2019).



\bibitem{Heemskerk:2009pn} 
  I.~Heemskerk, J.~Penedones, J.~Polchinski and J.~Sully,
  ``Holography from Conformal Field Theory,''
  JHEP {\bf 0910}, 079 (2009)
  [arXiv:0907.0151 [hep-th]].



\bibitem{Faulkner:2013ica} 
  T.~Faulkner, M.~Guica, T.~Hartman, R.~C.~Myers and M.~Van Raamsdonk,
  ``Gravitation from Entanglement in Holographic CFTs,''
  JHEP {\bf 1403}, 051 (2014)
  [arXiv:1312.7856 [hep-th]].


\bibitem{Shenker:2013pqa} 
  S.~H.~Shenker and D.~Stanford,
  ``Black holes and the butterfly effect,''
  JHEP {\bf 1403}, 067 (2014)
  [arXiv:1306.0622 [hep-th]].

\bibitem{Maldacena:2015waa} 
  J.~Maldacena, S.~H.~Shenker and D.~Stanford,
  ``A bound on chaos,''
  JHEP {\bf 1608}, 106 (2016)
  [arXiv:1503.01409 [hep-th]].


\bibitem{Kitaev-talk}
A.~Kitaev, 
talk given at Fundamental Physics Symposium, Nov.~2014.


\bibitem{Hawking:1982dh}
  S.~W.~Hawking and D.~N.~Page,
  ``Thermodynamics of Black Holes in anti-De Sitter Space,''
  Commun.\ Math.\ Phys.\  {\bf 87} (1983) 577.


\bibitem{Klebanov:1999tb}
  I.~R.~Klebanov and E.~Witten,
  ``AdS / CFT correspondence and symmetry breaking,''
  Nucl.\ Phys.\ B {\bf 556} (1999) 89
  [hep-th/9905104].


\bibitem{Optics}
K.~Sharma, ``Optics: principles and applications'', Academic Press, 2006.


\bibitem{Nambu:2012wa}
  Y.~Nambu,
  ``Wave Optics and Image Formation in Gravitational Lensing,''
  J.\ Phys.\ Conf.\ Ser.\  {\bf 410} (2013) 012036
  [arXiv:1207.6846 [gr-qc]].

\bibitem{Kanai:2013rga} 
  K.~i.~Kanai and Y.~Nambu,
  ``Viewing Black Holes by Waves,''
  Class.\ Quant.\ Grav.\  {\bf 30}, 175002 (2013)
  [arXiv:1303.5520 [gr-qc]].

\bibitem{Nambu:2015aea}
  Y.~Nambu and S.~Noda,
  ``Wave Optics in Black Hole Spacetimes: Schwarzschild Case,''
  Class.\ Quant.\ Grav.\  {\bf 33} (2016) 075011
  [arXiv:1502.05468 [gr-qc]].


\bibitem{Cardoso:2001hn}
  V.~Cardoso and J.~P.~S.~Lemos,
  ``Scalar, electromagnetic and Weyl perturbations of BTZ black holes: Quasinormal modes,''
  Phys.\ Rev.\ D {\bf 63} (2001) 124015
  [gr-qc/0101052].

\bibitem{Natsuume:2014sfa}
  M.~Natsuume,
  ``AdS/CFT Duality User Guide,''
  Lect.\ Notes Phys.\  {\bf 903} (2015) pp.1
  [arXiv:1409.3575 [hep-th]].

\bibitem{Ferrari:1984zz}
  V.~Ferrari and B.~Mashhoon,
  ``New approach to the quasinormal modes of a black hole,''
  Phys.\ Rev.\ D {\bf 30} (1984) 295.

\bibitem{Festuccia:2008zx}
  G.~Festuccia and H.~Liu,
  ``A Bohr-Sommerfeld quantization formula for quasinormal frequencies of AdS black holes,''
  Adv.\ Sci.\ Lett.\  {\bf 2} (2009) 221
  [arXiv:0811.1033 [gr-qc]].

\bibitem{Berti:2009wx}
  E.~Berti, V.~Cardoso and P.~Pani,
  ``Breit-Wigner resonances and the quasinormal modes of anti-de Sitter black holes,''
  Phys.\ Rev.\ D {\bf 79} (2009) 101501
  [arXiv:0903.5311 [gr-qc]].

\bibitem{Berti:2009kk}
  E.~Berti, V.~Cardoso and A.~O.~Starinets,
  ``Quasinormal modes of black holes and black branes,''
  Class.\ Quant.\ Grav.\  {\bf 26} (2009) 163001
  [arXiv:0905.2975 [gr-qc]].

\bibitem{Dias:2012tq}
  O.~J.~C.~Dias, G.~T.~Horowitz, D.~Marolf and J.~E.~Santos,
  ``On the Nonlinear Stability of Asymptotically Anti-de Sitter Solutions,''
  Class.\ Quant.\ Grav.\  {\bf 29} (2012) 235019
  [arXiv:1208.5772 [gr-qc]].



\bibitem{Gubser:2008px}
  S.~S.~Gubser,
  ``Breaking an Abelian gauge symmetry near a black hole horizon,''
  Phys.\ Rev.\ D {\bf 78} (2008) 065034
  [arXiv:0801.2977 [hep-th]].

\bibitem{Hartnoll:2008vx}
  S.~A.~Hartnoll, C.~P.~Herzog and G.~T.~Horowitz,
  ``Building a Holographic Superconductor,''
  Phys.\ Rev.\ Lett.\  {\bf 101} (2008) 031601
  [arXiv:0803.3295 [hep-th]].

\bibitem{Hartnoll:2008kx}
  S.~A.~Hartnoll, C.~P.~Herzog and G.~T.~Horowitz,
  ``Holographic Superconductors,''
  JHEP {\bf 0812} (2008) 015
  [arXiv:0810.1563 [hep-th]].

\bibitem{Sachdev:2011wg}
  S.~Sachdev,
  ``What can gauge-gravity duality teach us about condensed matter physics?,''
  Ann.\ Rev.\ Condensed Matter Phys.\  {\bf 3} (2012) 9
  [arXiv:1108.1197 [cond-mat.str-el]].

\bibitem{Fujita:2014mqa}
  M.~Fujita, S.~Harrison, A.~Karch, R.~Meyer and N.~M.~Paquette,
  ``Towards a Holographic Bose-Hubbard Model,''
  JHEP {\bf 1504} (2015) 068
  [arXiv:1411.7899 [hep-th]].

\bibitem{BHMOTOC}
H.~Shen, P.~Zhang, R.~Fan and H.~Zhai, 
``Out-of-time-order correlation at a quantum phase transition''
Phys.\ Rev.\ B {\bf 96} 054503 (2017).




\end{thebibliography}
\end{document}